\documentclass{siamart171218}

\usepackage[margin=1.0in]{geometry}
\usepackage[utf8]{inputenc}

\usepackage{graphicx}
\usepackage[version=4]{mhchem}
\usepackage{siunitx} 
\usepackage{longtable,tabularx}
\usepackage{xcolor}
\usepackage{tablefootnote}
\usepackage{siunitx}

\usepackage{subfigure}			
\usepackage{subfigmat}			

\usepackage{comment}

\usepackage{multirow}

\newcommand{\coloring}[1]{\textcolor{black}{#1}} 
\newcommand{\X}{$\times$}

\newcommand{\commentout}[1]{{}} 

\renewcommand{\theequation}{\arabic{section}.\arabic{equation}}

\newcommand{\norm}[1]{\left\|#1\right\|}

\newcommand {\real} {I\!\!R}

\title{PIFE-PIC: Parallel Immersed-Finite-Element Particle-in-Cell
for 3-D Kinetic Simulations of Plasma-Material Interactions%
\thanks{%
Submitted to the editors \today.
\funding{This work was financially supported in part by
NASA-Missouri Space Grant Consortium through NASA-EPSCoR-Missouri,
and NSF through grants DMS-2005272 and OAC-1919789. 
D. Lund was also supported in part by NASA-Missouri Space Grant Consortium Scholarships.
} 
} 
} 

\author{
Daoru Han\footnotemark[2]%
\thanks{Department of Mechanical and Aerospace Engineering,
Missouri University of Science and Technology, Rolla, MO 65409
(\email{handao@mst.edu},\email{dclgzb@mst.edu}).}
\and
Xiaoming He%
\thanks{Department of Mathematics and Statistics,
Missouri University of Science and Technology, Rolla, MO 65409 (\email{hex@mst.edu})}
\and
David Lund\footnotemark[2]%
\and
Xu Zhang%
\thanks{Department of Mathematics, Oklahoma State University,
Stillwater, OK 74078 (\email{xzhang@okstate.edu})}}

\begin{document}
\newcommand{\BibTeX}{{\scshape Bib}\TeX\xspace}
\maketitle
\begin{abstract}
This paper presents a recently developed particle simulation code package PIFE-PIC,
which is a novel three-dimensional (3-D)
Parallel Immersed-Finite-Element (IFE) Particle-in-Cell (PIC) simulation model
for particle simulations of plasma-material interactions.
This framework is based on the recently developed non-homogeneous
electrostatic IFE-PIC algorithm,
which is designed to handle complex plasma-material interface conditions
associated with irregular geometries using a Cartesian-mesh-based PIC.
Three-dimensional domain decomposition is utilized
for both the electrostatic field solver with IFE
and the particle operations in PIC
to distribute the computation among multiple processors.
A simulation of the orbital-motion-limited (OML) sheath of a dielectric sphere
immersed in a stationary plasma is carried out to validate PIFE-PIC
and profile the parallel performance of the code package.
Furthermore, a large-scale simulation
of plasma charging at a lunar crater
containing 2 million PIC cells (10 million FE/IFE cells)
and about 520 million particles,
running for 20,000 PIC steps in about 109 wall-clock hours,
is presented to demonstrate the high-performance computing capability of PIFE-PIC.

\end{abstract}

\begin{keywords}
	immersed-finite-element,
	particle-in-cell,
    parallel domain decomposition,
	plasma-material interactions
\end{keywords}

\begin{AMS}
	35R05, 65N30, 65Y05
\end{AMS}

\section{Introduction}
\label{sec:intro}
\phantomsection

Particle modeling of plasma dynamics has emerged as one of the most
appropriate algorithms for first-principle-based modeling
of many plasma-material interaction (PMI)
problems.
One of the fundamental phenomena in plasma-material interactions is surface charging.
When an object is immersed in a plasma,
its surface will collect charge from the plasma
until it reaches an equilibrium surface potential determined by the current balance condition.
Many plasma-material interaction problems involve multiple objects with complex geometries,
therefore the interface conditions between the plasma and object need to be accurately resolved.

Being one of the most popular kinetic methods for collisionless plasma simulations,
the Particle-in-Cell (PIC) method \cite{1991BirdsallLangdon}
models the charged particles as macro-particles and tracks the motions of particles
in the electrostatic/electromagnetic field.
The electric potential in a PIC simulation domain
is governed by the second-order elliptic partial differential equations (PDEs)
with discontinuous dielectric coefficients
and non-homogeneous flux jumps across the material surface interface.
Numerical methods based on structured meshes, especially Cartesian meshes,
are particularly desirable in these simulations
because they enable efficient particle tracking and save computing time
in particle-mesh interactions.

The immersed-finite-element (IFE) method is a finite element method (FEM)
for solving interface problems using interface-independent meshes
such as Cartesian meshes.
The main idea of IFE is to adjust approximating functions locally
to accommodate the physical interface conditions.
An IFE method can achieve optimal convergence on an interface-independent mesh
with the number and location of the degrees-of-freedom isomorphic
to the standard FEM on the same mesh.
The first IFE method was introduced by Z. Li in \cite{1998Li}
for solving one-dimensional (1-D) elliptic interface problems with piecewise linear polynomials.
Since then, the IFE method has been extended to higher-order approximations
\cite{2009AdjeridLin, 2006CampLinLinSun, 2017CaoZhangZhang, 2016GuzmanSanchezSarkis2},
higher-dimensional elliptic interface problems
\cite{2020GuoLin2, 2020GuoZhang,2003LiLinWu, 2019HeZhang, 2015LinLinZhang,
2019LinSheenZhang, 2010VallaghePapadopoulo},
and other interface problems with other PDE models
\cite{2019AdjeridChaabaneLinYue,2020AdjeridLinZhuang,2019AdjeridMoon,2020GuoLinLin,
2013HeLinLinZhang,2013LinSheenZhang}.

Over the past decade, the IFE method has been successfully used
together with PIC in plasma particle simulations
\cite{2018BaiCaoChuZhang, 2016CaoChuZhangZhang,
2015JianChuCaoCaoHeXia,2005KafafyLinLinWang, 2005KafafyWangLin}.
Recently, a non-homogeneous IFE-PIC algorithm has been developed
for particle simulations of plasma-material interactions with complex geometries
while maintaining the computational speed of the Cartesian-mesh-based PIC
\cite{Chu_ijnam2017_software, Han_usc_dissertation2015, Han_ieee2016_charging,
Han_ife_jcp2016,CLu_JWan_YCao_XMHe_1,CLu_ZYang_JBai_YCao_XMHe_1}.
To the best of our knowledge, most existing IFE-PIC algorithms are serial.
The non-parallel algorithms have limitations in their capability
to handle large-scale particle simulations
and their efficiency in using multiple processors at the algorithm level.
For a typical large-scale 3-D PIC simulation,
millions to billions of particles are tracked in the computation domain
that contains millions of elements.
With the availability of multi-processor computational facilities,
the call for parallel IFE-PIC algorithms is urgent.

The goal of this paper is to develop and test a new Parallel IFE-PIC package
for particle simulations of electrostatic plasma-material interactions,
namely, PIFE-PIC.
We utilize a 3-D domain decomposition technique
for both \emph{field-solve} and \emph{particle-push} procedures of the PIC model.
The computations are distributed into multiple subdomains
which can be handled independently by multiple processors.
The key is how to efficiently exchange the information between these subdomains.
In this work, neighboring subdomains have a small overlapping (``guard cells'') region
which will be used as a common region to interchange the PDE solutions and the particle data.
Extensive numerical experiments show that
our PIFE-PIC scheme significantly outperforms the serial IFE-PIC scheme.
Although it maintains a similar accuracy as the serial IFE-PIC computational scheme,
the high parallel performance dramatically reduces the computational time
for problems of practical interests.
Hence, large-scale kinetic simulations of plasma-material interactions
can be carried out much more efficiently.

The rest of this paper is organized as follows.
In Section \ref{sec:pifepic}, we describe the details of 3-D domain decomposition
for both IFE (field-solve) and PIC (particle-push) procedures of PIFE-PIC.
In Section \ref{sec:validation}, we present a code validation
using a 3-D sheath problem
of a dielectric sphere immersed in a stationary plasma.
Section \ref{sec:scaling} presents a parallel efficiency test of
the PIFE-PIC code for strong scaling.
Section \ref{sec:lunar} presents an application of PIFE-PIC
to simulations of lunar surface charging at a crater.
Finally, a summary and conclusion are given in Section \ref{sec:conclusion}.

\section{Parallel IFE-PIC Algorithms}
\label{sec:pifepic}
\phantomsection

\subsection{Overview of PIC and IFE-PIC}

\begin{figure}[ht!]
\centering
{\includegraphics[width=0.5\textwidth]{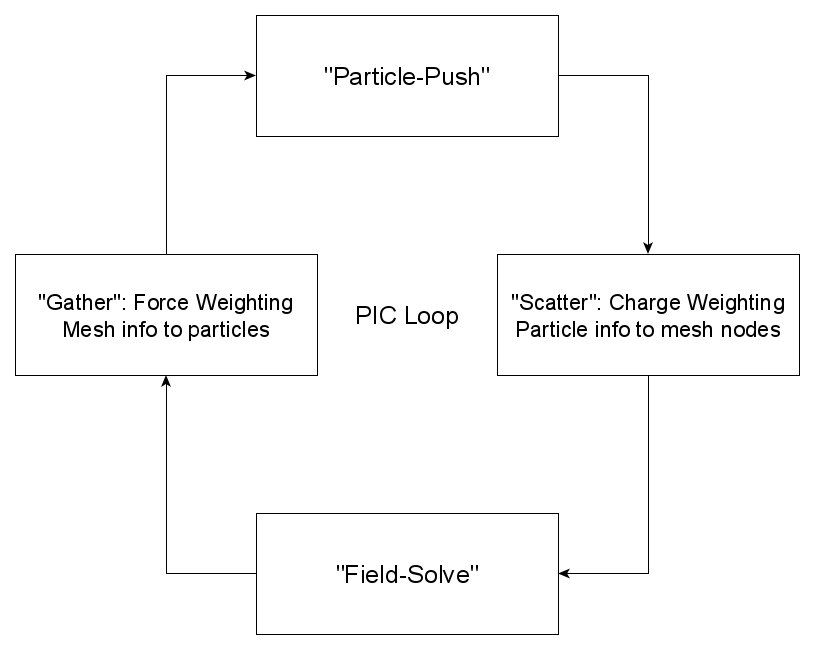}}
\caption{Four essential steps in a PIC loop.}
\label{fig:picloop}
\end{figure}

PIC is a widely-used kinetic particle simulation method for plasma dynamics
\cite{Birdsall_Langdon_2005,Hockney_Eastwood_1988}.
In PIC, charged particles of plasma species are represented
by a number of simulation particles
(also referred to as macro-particles or super-particles)
distributed ``freely'' in the entire computation domain,
while the field quantities such as electric potential are discretized on a mesh
(thus the name ``particle-in-cell'').
The kernel of PIC method is the ``PIC loop''
which includes four essential steps: scatter, field-solve, gather, and particle-push
(Figure \ref{fig:picloop}).
Within a PIC loop, quantities carried by the simulation particles
are weighted onto the mesh nodes (``scatter'') to form the right-hand side (RHS) term
of the PDE for the solution of the electrostatic/electromagnetic field (``field-solve''),
which is in turn interpolated at particle positions (``gather'')
to update the velocity and position of the particles (``particle-push'').
Such data exchange between particles and field quantities
will iterate for a desired number of steps
(or till a convergence criterion is met)
to obtain the self-consistent solution of both particles and fields.

\begin{figure}[ht!]
\centering
\begin{subfigmatrix}{2} 
\subfigure[Computation domain including the interface]%
{\includegraphics[width=0.35\textwidth]{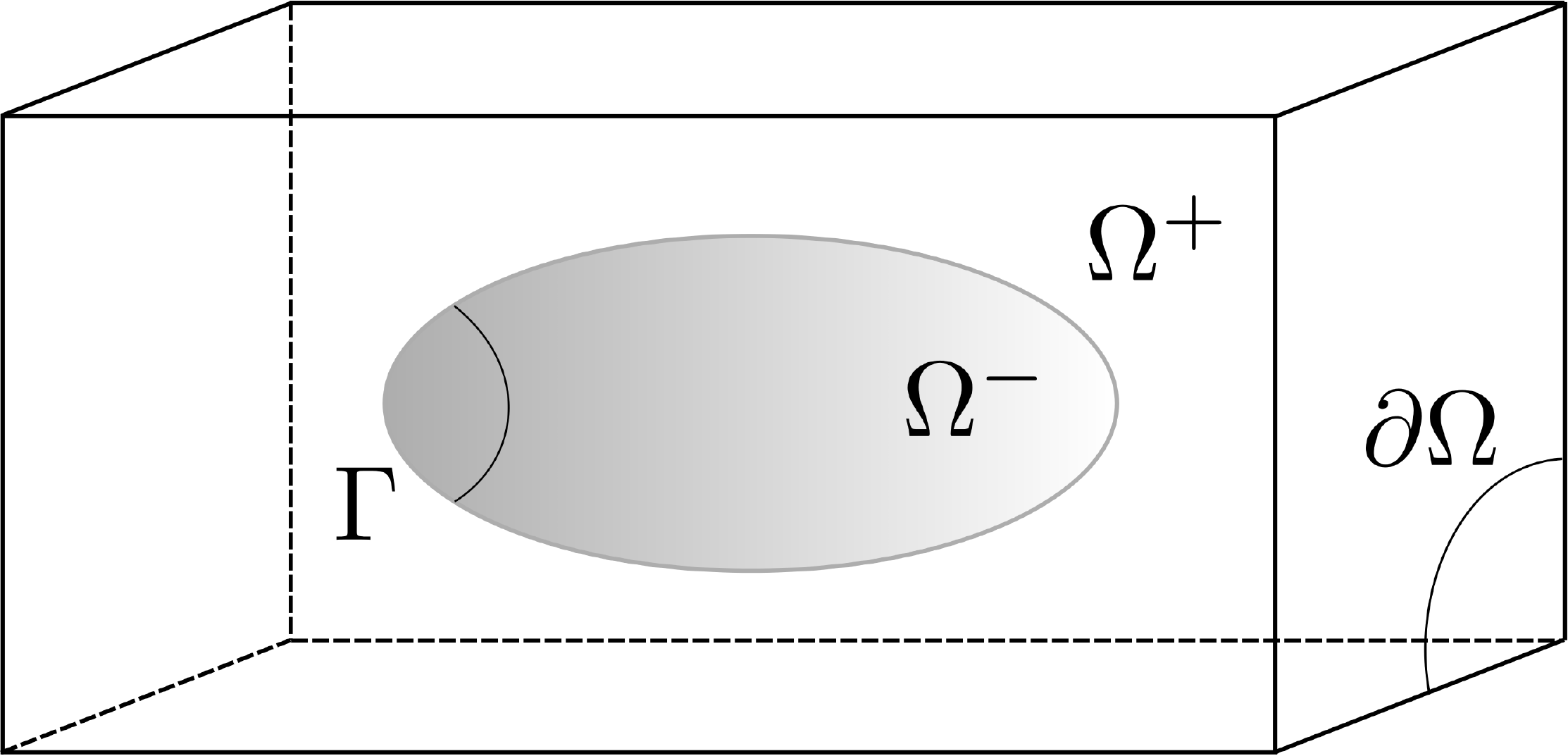}\label{fig:domain}}
\subfigure[Electric flux jump across the interface]%
{\includegraphics[width=0.4\textwidth]{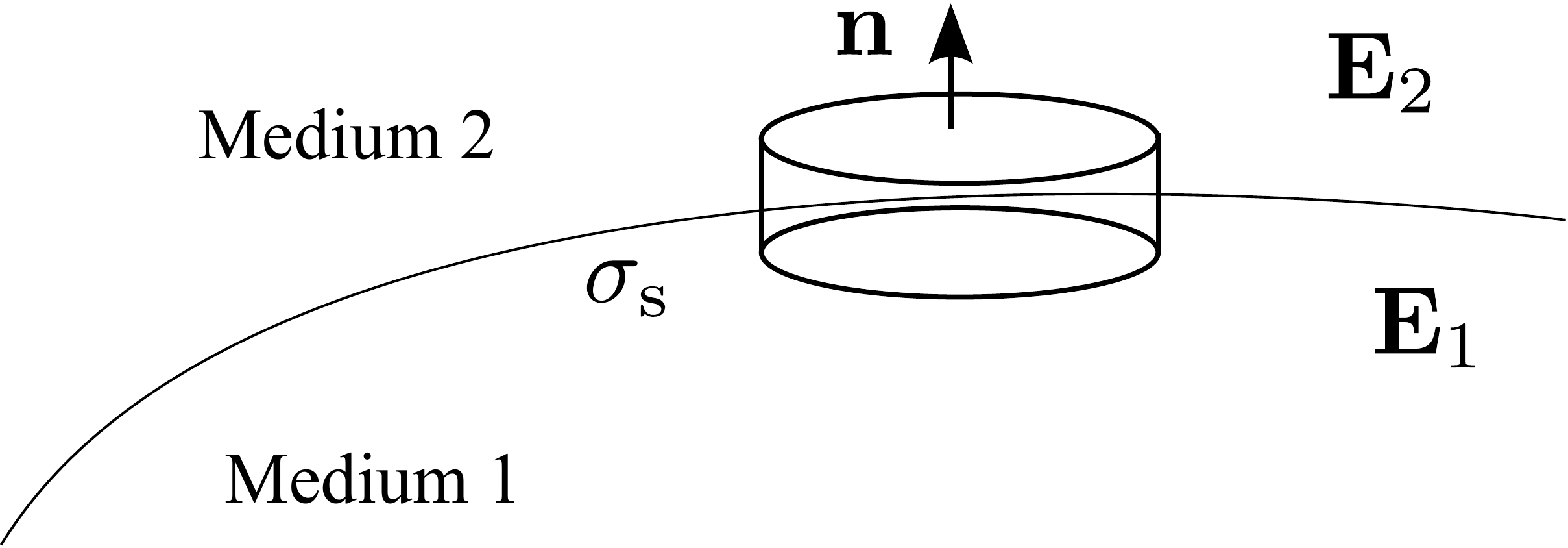}\label{fig:interface}}
\end{subfigmatrix}
\caption{A sketch of the interface problem and interface condition.}
\label{fig:sketch}
\end{figure}

For problems of plasma-material interactions,
the mathematical model
is an interface problem including
the electrostatic/electromagnetic field problem
in a self-consistent solution to the corresponding plasma dynamics problem
(Figure \ref{fig:domain}),
together with the appropriate interface conditions
between the plasma region and the material region (Figure \ref{fig:interface}).
For electrostatic problems presented in this work,  
we consider the following boundary value problem of the elliptic equation
that governs the distribution of the electric potential $\phi$
\cite{JDJackson_book_1}:
\begin{eqnarray}
-\nabla \cdot \big( \varepsilon \nabla \phi(X) \big) &=&
\rho(X),~~X=(x,y,z) \in {\Omega^- \cup \Omega^+},
\label{bvp_pde_nonhomogeneous flux jump} \\
\phi(X) &=& g(X),~~ X \in {\Gamma_D},
\label{eq:bvp_bc_nonhomogeneous flux jump} \\
 \frac{\partial \phi(X)}{\partial\textbf{n}_{\Gamma_N}} &=& p(X),~~ X \in {\Gamma_N}.
\label{eq:bvp_bc2_nonhomogeneous flux jump}
\end{eqnarray}
Here, $\Omega\in \real^3$ is assumed to be a cuboidal domain,
which is divided into two subdomains $\Omega^+$ and $\Omega^-$
by an interface surface $\Gamma$ such that
$\overline{\Omega}= \overline{\Omega^-} \cup \overline{\Omega^+} \cup \Gamma$.
$\Gamma_D$ and $\Gamma_N$ are the Dirichlet and Neumann boundaries
such that $\partial\Omega = \Gamma_D \cup \Gamma_N$.
The vector $\textbf{n}_{\Gamma_N}$ is the unit outward normal of $\Gamma_N$.
See the sketch in Figure \ref{fig:domain}.
The functions $\rho$, $g$, and $p$ are
the source term, Dirichlet boundary function, and Neumann boundary function, respectively.
The electric field $\mathbf{E} = -\nabla \phi(X)$ is discontinuous
across the interface $\Gamma$ with the following jump conditions imposed:
\begin{eqnarray}
\left. \left[\phi(X) \right] \right|_\Gamma &=& 0,
\label{eq:bvp_int_1_nonhomogeneous flux jump} \\
\left. \left[
\varepsilon \frac{\partial \phi(X)}{\partial\textbf{n}_{\Gamma}}
\right]
\right|_{\Gamma}
&=& q(X),
\label{eq:bvp_int_2_nonhomogeneous flux jump}
\end{eqnarray}
where the jump $[\cdot]_\Gamma$ is defined by
$[w(X)]_\Gamma := w^+(X)|_\Gamma - w^-(X)|_\Gamma$.
The vector $\textbf{n}_{\Gamma}$ is the unit normal of $\Gamma$
pointing from $\Omega^-$ to $\Omega^+$.
The material-dependent coefficient $\varepsilon{(X)}$ is discontinuous across the interface.
Without loss of generality, we assume it is a piecewise constant function defined by
\begin{eqnarray*}
\varepsilon(X) = \left\{ \begin{array} {ll}
\varepsilon^-, ~~X \in \Omega^-, \\
\varepsilon^+, ~~X \in \Omega^+,
\end{array} \right.
\end{eqnarray*}
where $\min(\varepsilon^+,\varepsilon^-) >0$.

\begin{figure}[ht!]
\centering
\begin{subfigmatrix}{2} 
\subfigure[Unstructured body-fitting finite element (FE) mesh]%
{\label{fig:ifepic:mesh:a}\includegraphics[width=0.45\textwidth]{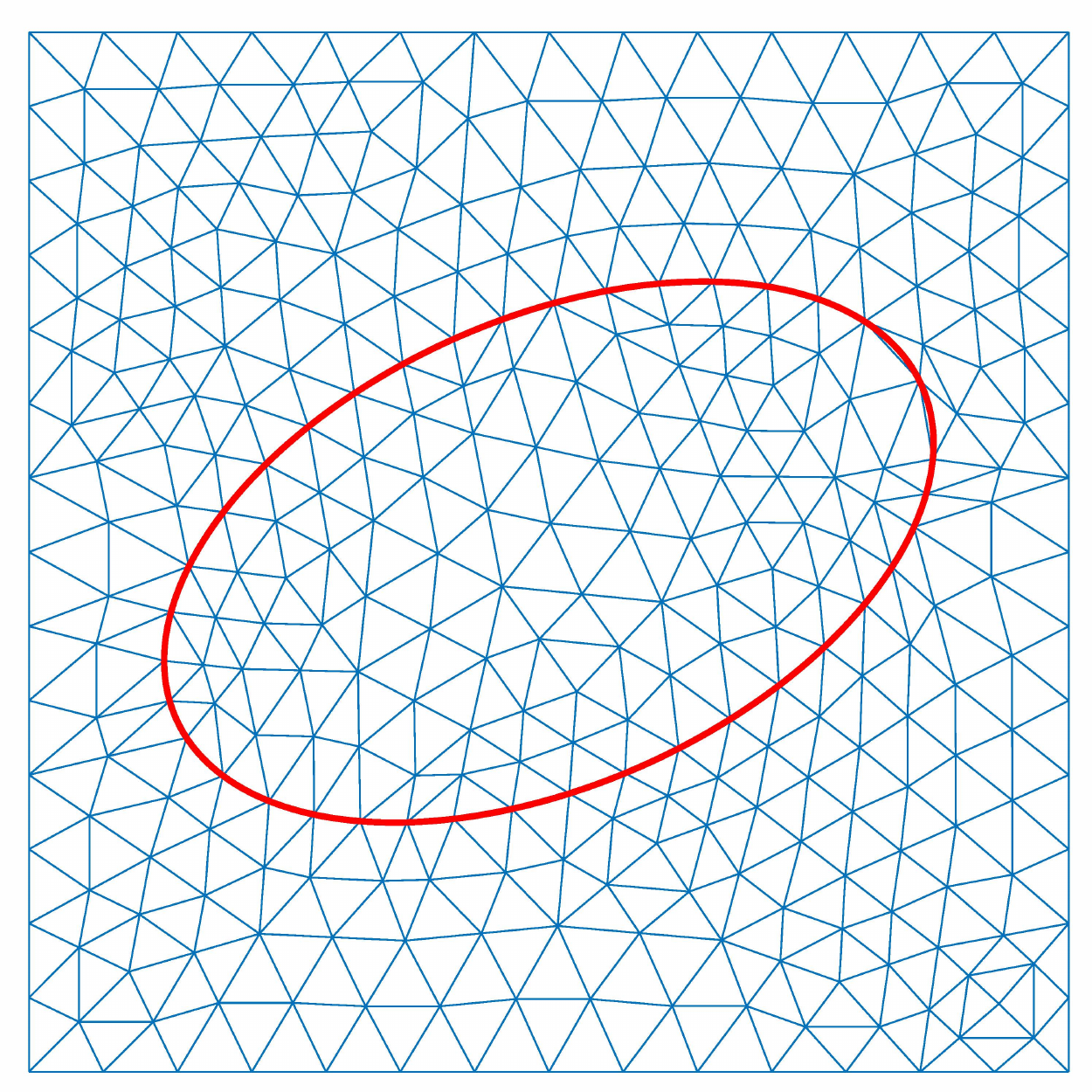}}
\subfigure[Structured immersed-finite-element (IFE) mesh based on Cartesian mesh]%
{\label{fig:ifepic:mesh:b}\includegraphics[width=0.45\textwidth]{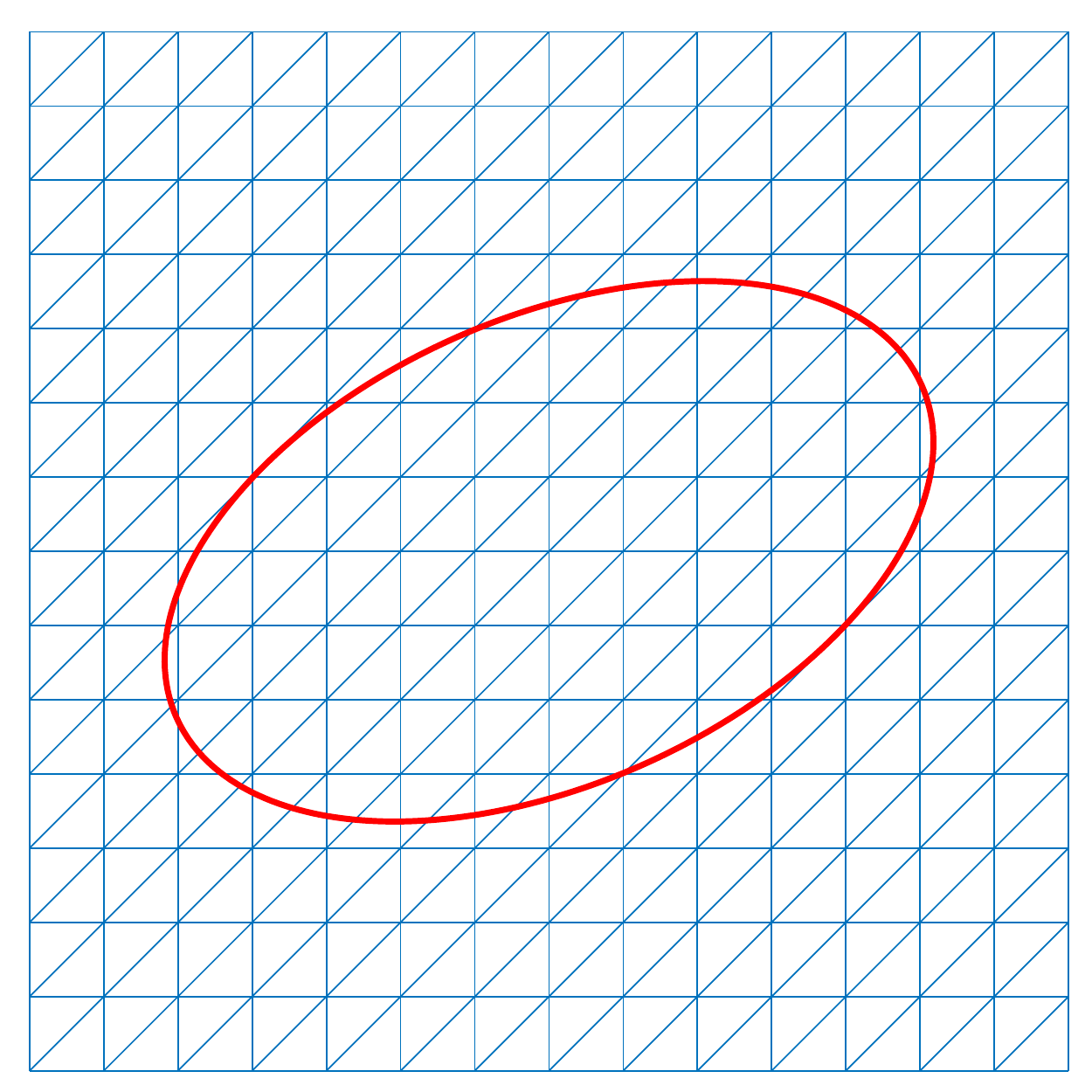}}
\end{subfigmatrix}
\caption{Illustration of traditional body-fitting finite-element mesh and
novel structured immersed-finite-element mesh.}
\label{fig:ifepic:mesh}
\end{figure}

\begin{figure}[ht!]
\centering
{\includegraphics[width=0.8\textwidth]{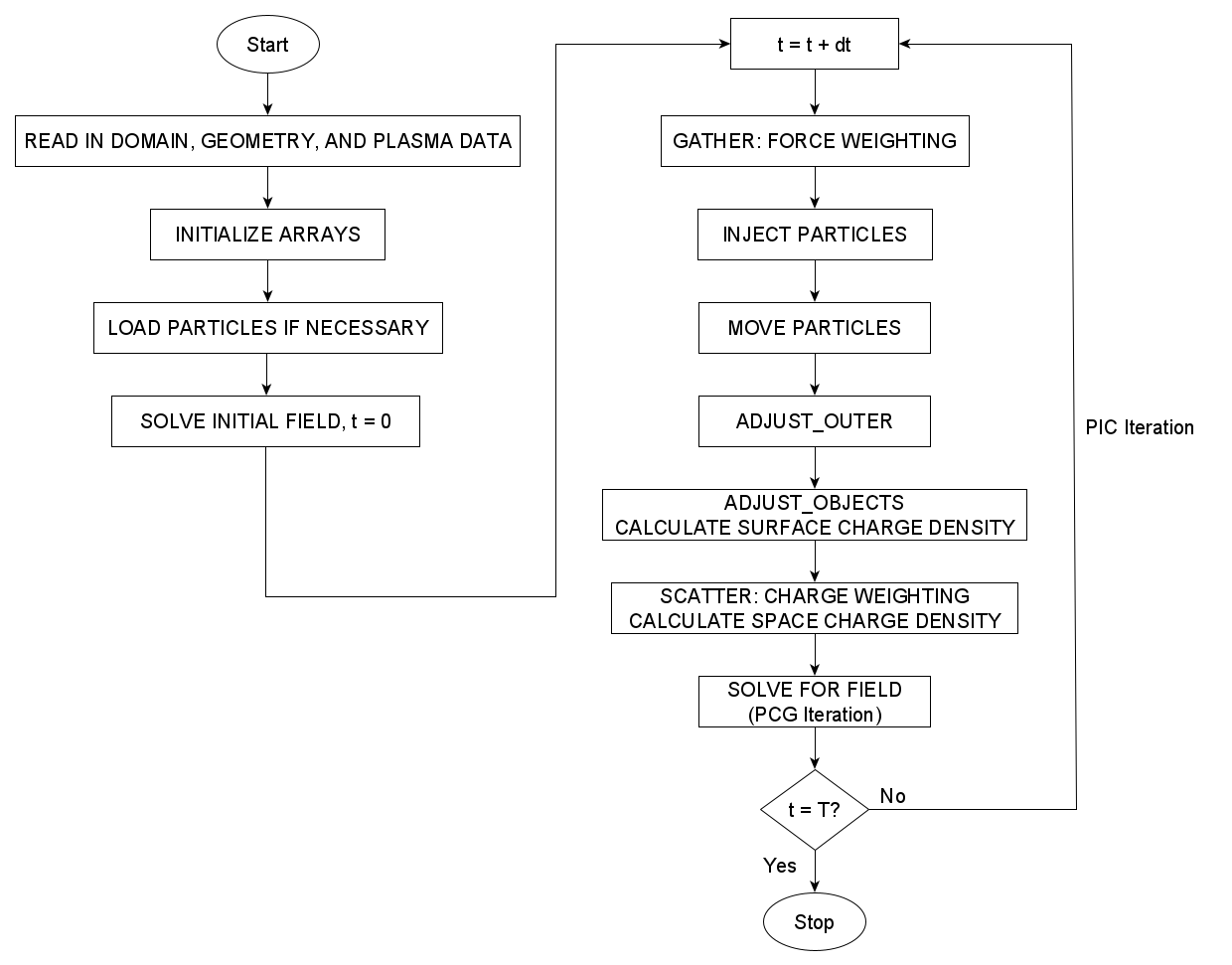}}
\caption{Flowchart of serial IFE-PIC.}
\label{fig:flowchart:serial}
\end{figure}

In many applications of scientific and engineering interest,
the shape of the interface $\Gamma$ is usually non-trivial.
Traditionally, when solving field problems involving complex-shaped objects,
an unstructured body-fitting mesh is employed to improve accuracy
(Figure \ref{fig:ifepic:mesh:a}).
However, a structured mesh, such as Cartesian mesh (Figure \ref{fig:ifepic:mesh:b}),
is more advantageous in kinetic PIC modeling
of plasma dynamics from the perspective of computing speed and efficiency,
although, it has been limited to problems with relatively simple geometries
due to accuracy considerations inherited from finite-difference-based schemes.
To solve this dilemma while taking into account both accuracy and efficiency,
the immersed-finite-element particle-in-cell (IFE-PIC) method
was developed
to handle complex interface conditions associated with irregular geometries
while maintaining the computational speed of the Cartesian-mesh-based PIC.
The detailed IFE formulation and IFE-PIC steps
are archived in \cite{Han_ife_jcp2016}
and the flowchart of the serial IFE-PIC algorithm
is shown in Figure \ref{fig:flowchart:serial}.
Over the past few years, the IFE-PIC method has matured to successfully model
plasma dynamics problems arising from many space applications,
such as ion thruster grid optics
\cite{RK_AIAA_2005_3691_WholeSubscaleIonOptics,RK_JPP_2007_WholeIonOptics},
ion propulsion plume-induced contamination
\cite{RK_IEEE_2006_Plume_HybridGrid,JW_IEEE_2006_PlumeParallel,
DH_AIAA_2013_3888_JPC},
charging of lunar and asteroidal surfaces
\cite{Han_usc_dissertation2015,Han_ife_jcp2016,Han_ieee2016_charging,
YCao_YChu_XMHe_TLin_1,
Han_jsr_2018_lunar,
Han_ieeetps2019_asteroid,
Will_ieeetps2019_asteroid},
and dust transport dynamics around small asteroids
\cite{Yu_AIAA2016_5447_asteroid}.


\subsection{3-D Domain Decomposition in PIFE-PIC}

\begin{figure}[ht!]
\centering
{\includegraphics[width=0.72\textwidth]{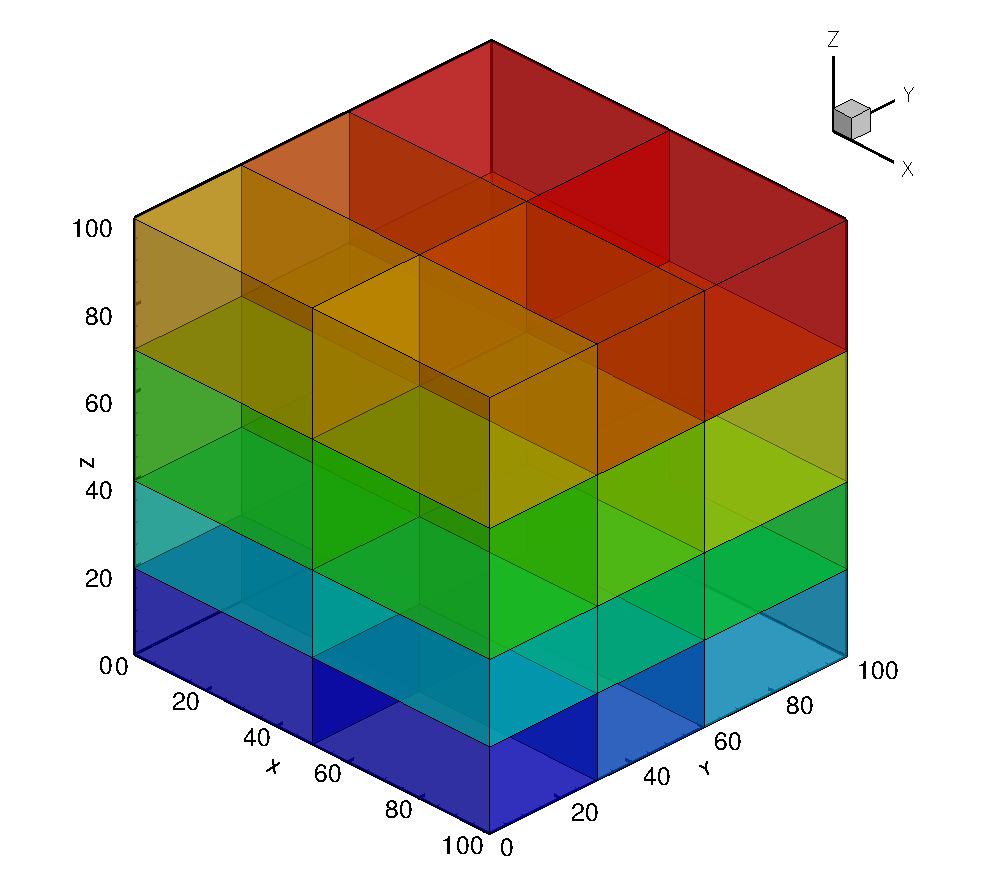}}
\caption{3-D domain decomposition for PIC blocks. Overlapping cells are not displayed.
In this example, the global domain is decomposed into 2$\times$3$\times$4 subdomains.
The blue-red color scale indicates the MPI rank of each subdomain.}
\label{fig:pifepic:blocks}
\end{figure}

\begin{figure}[ht!]
\centering
\begin{subfigmatrix}{3} 
\subfigure[Subdomains with two overlapping PIC cells in each dimension,
globally uniform PIC mesh.]
{\includegraphics[trim={4cm 0.5cm 9cm 0}, clip, width=0.25\textwidth]%
{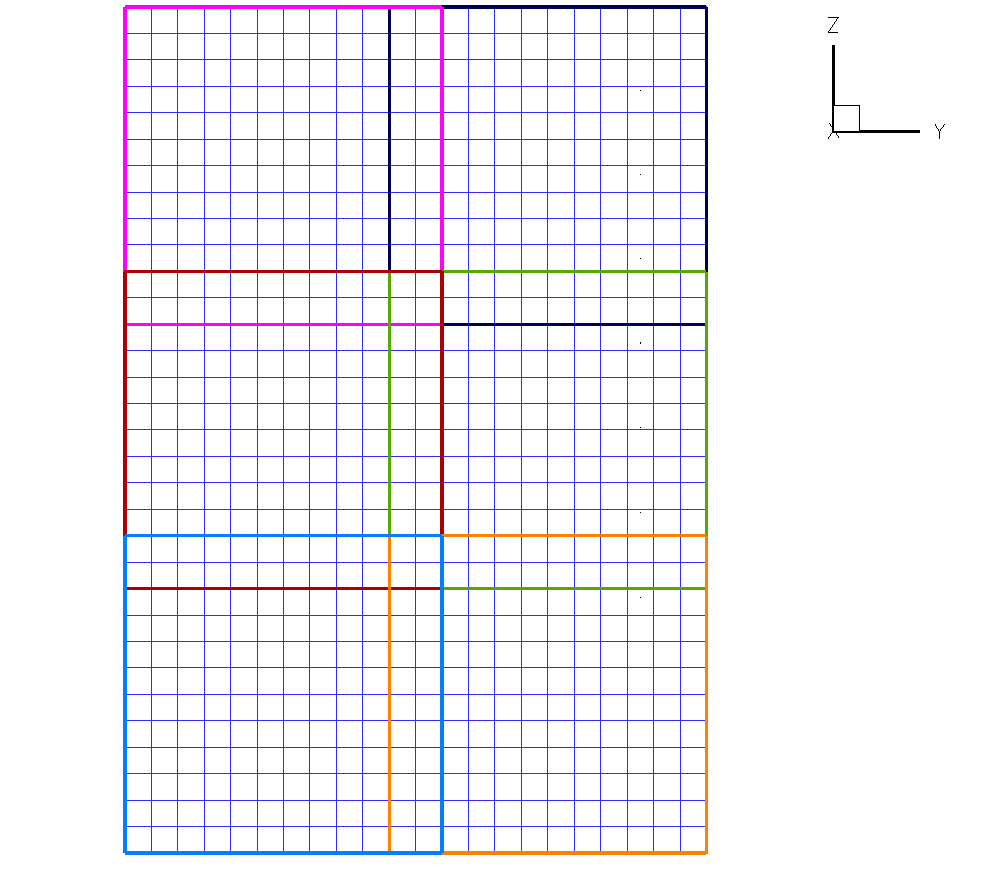}}
\subfigure[Subdomains with two overlapping PIC cells in each dimension,
globally uniform IFE mesh.]
{\includegraphics[trim={3cm 0 10cm 0}, clip, width=0.25\textwidth]%
{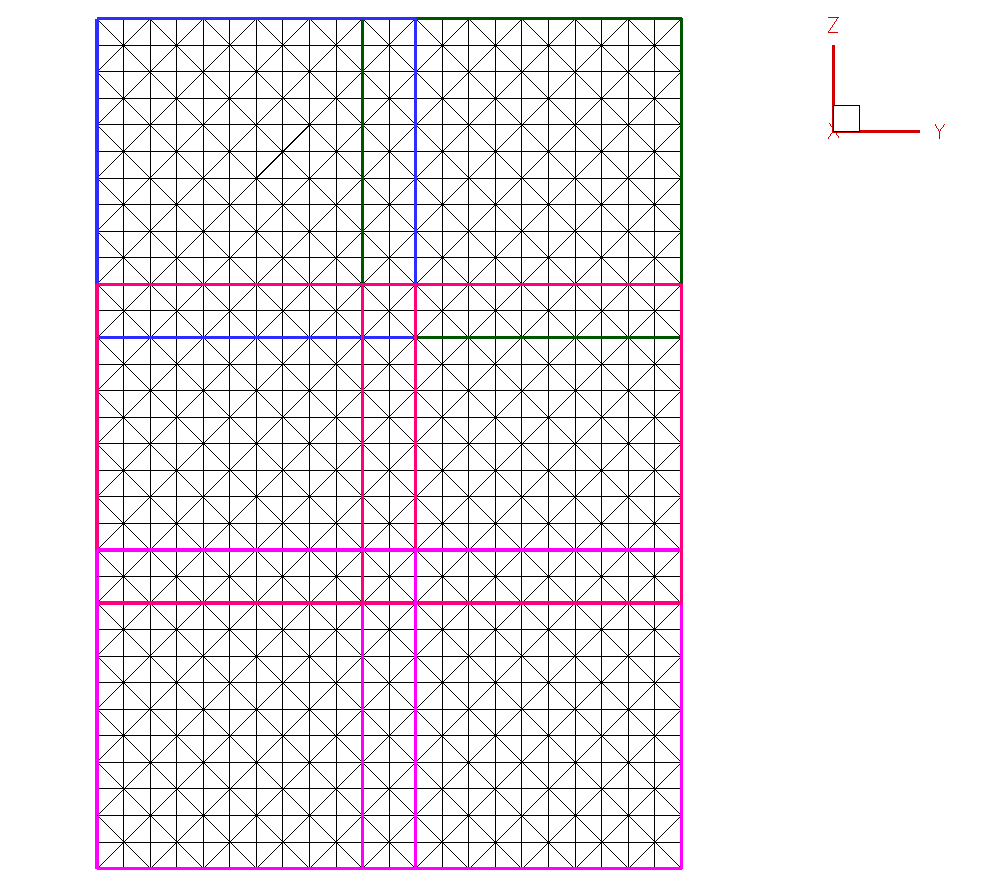}}
\subfigure[Subdomains with two overlapping PIC cells in each dimension,
globally non-uniform IFE mesh.]%
{\includegraphics[trim={3cm 0 10cm 0}, clip, width=0.25\textwidth]%
{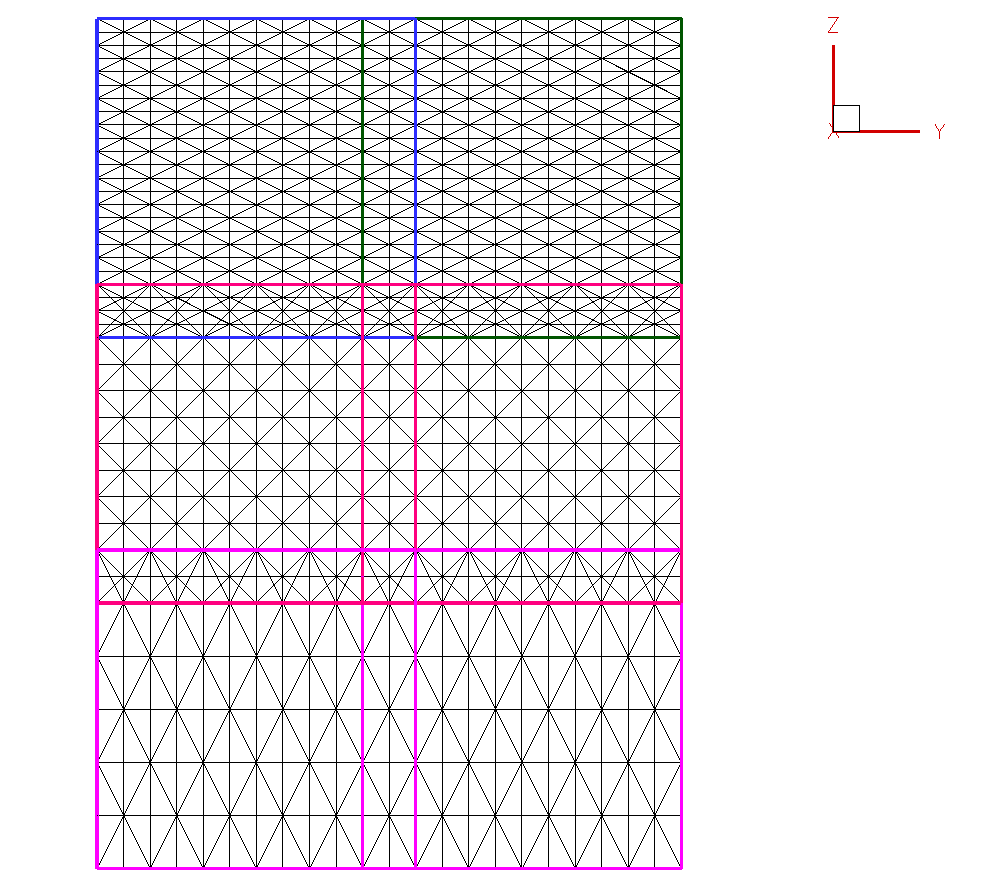}}
\end{subfigmatrix}
\caption{2-D projection showing the domain decomposition for PIC and IFE
with overlapping cells and different resolutions.}
\label{fig:2dmesh}
\end{figure}

\begin{figure}[ht!]
\centering
\begin{subfigmatrix}{2} 
\subfigure[A globally uniform IFE mesh.]%
{\includegraphics[trim={6cm 3cm 2cm 0}, clip, width=0.45\textwidth]%
{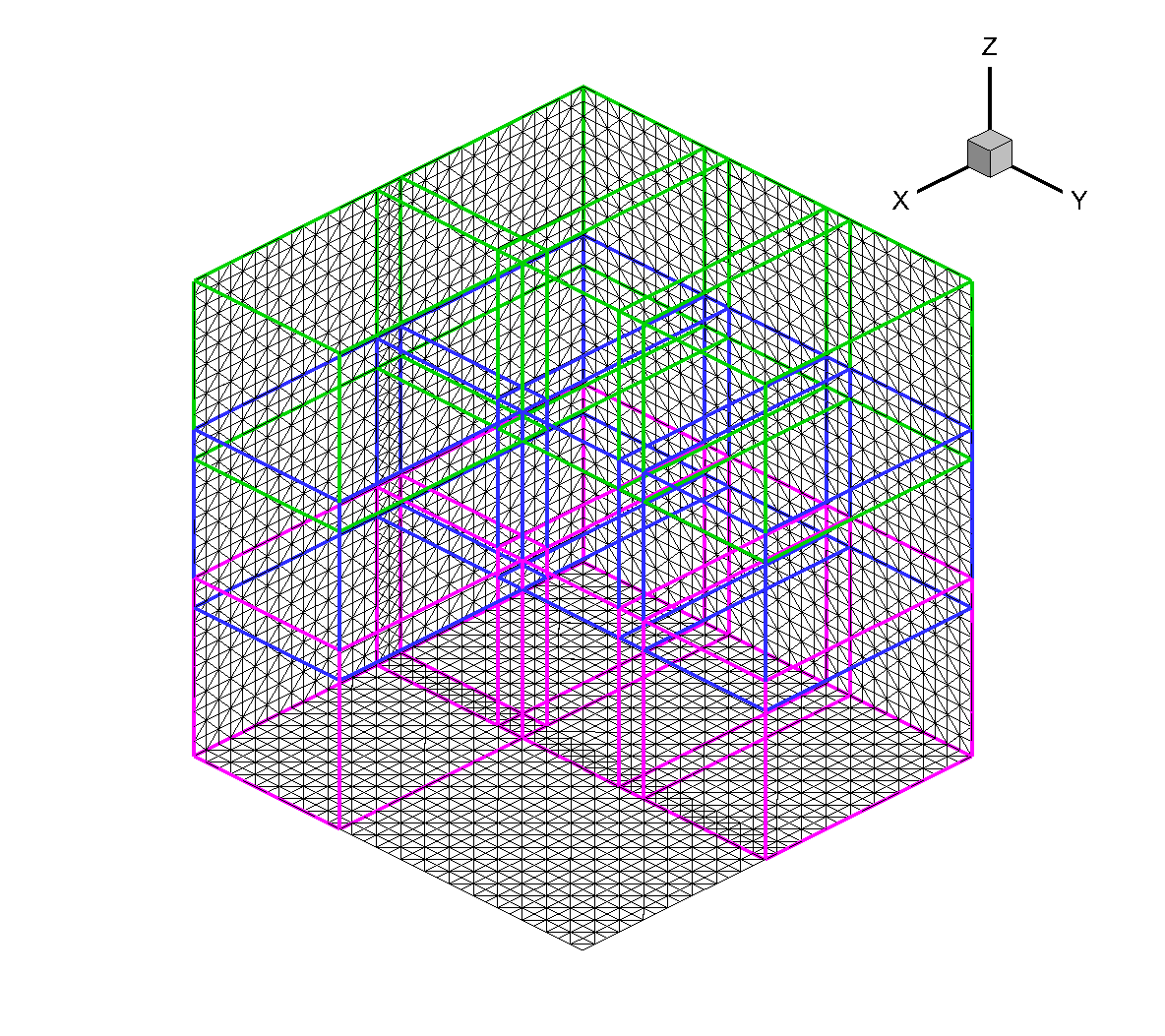}}
\subfigure[A globally non-uniform IFE mesh.]%
{\includegraphics[trim={6cm 3cm 2cm 0}, clip, width=0.45\textwidth]%
{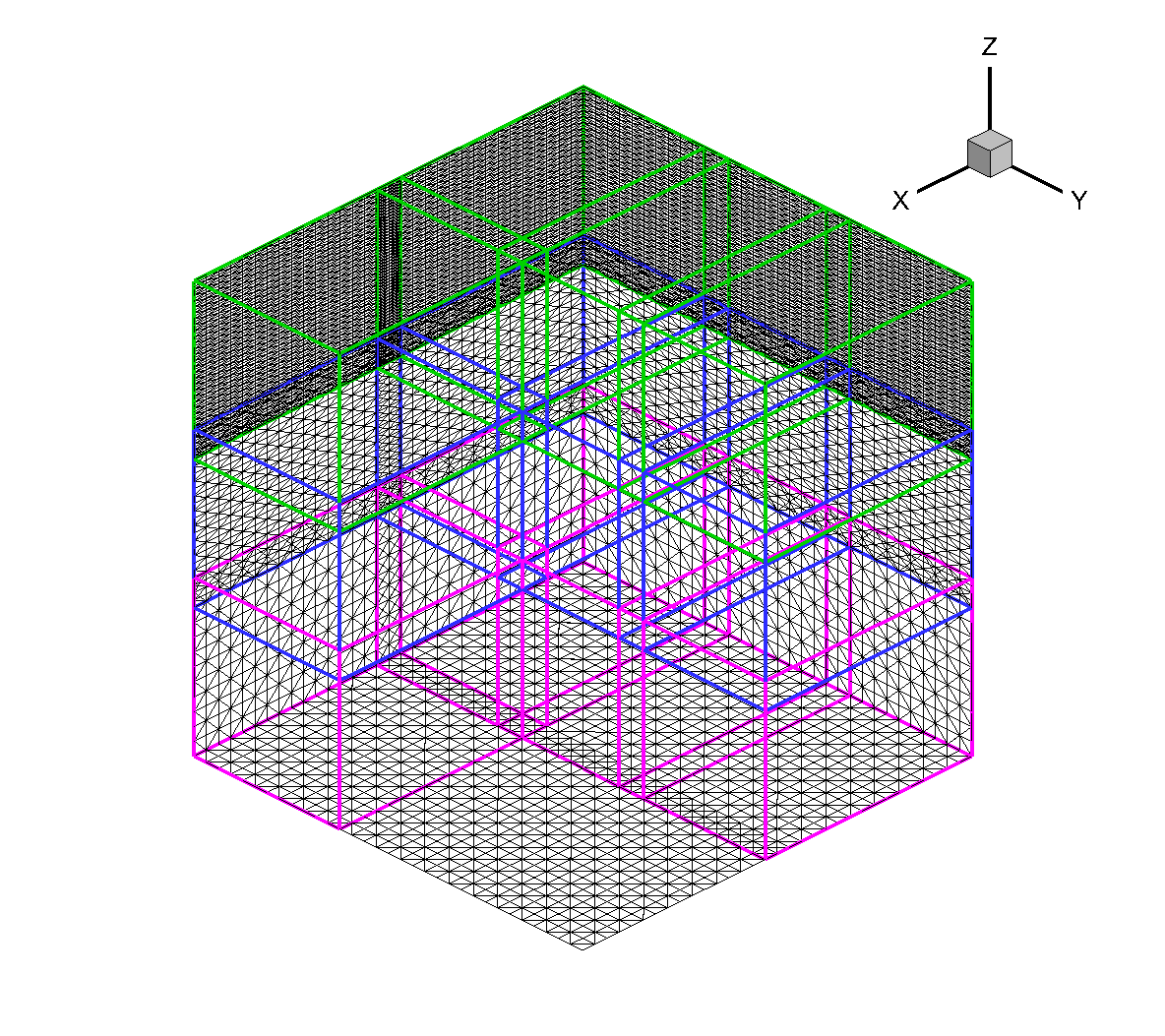}}
\end{subfigmatrix}
\caption{%
3-D view of globally uniform and non-uniform IFE meshes.
The IFE mesh for each subdomain is uniform (locally),
but could be non-uniform for different subdomains (globally).
}
\label{fig:3dmesh}
\end{figure}

In our proposed PIFE-PIC algorithm,
the 3-D computational domain is decomposed along each dimension
using the Message Passing Interface (MPI) architecture (Figure \ref{fig:pifepic:blocks}).
The domain is first decomposed into cuboid blocks with the same PIC mesh resolution.
Each subdomain is handled by a processor for both field-solve
and particle-push procedures of the PIC method.
Two overlapping PIC cells (``guard cells'') in each dimension are used in PIFE-PIC
(Figure \ref{fig:2dmesh}).
Therefore, the boundaries of each subdomain are either on the global boundary
or in the interior of its neighboring subdomains.
Local IFE mesh is then generated for each subdomain.
By virtue of the IFE formulation,
PIC and IFE can use different mesh resolutions.
In PIFE-PIC, PIC mesh is globally uniform.
However, IFE mesh could be globally non-uniform
but still locally uniform within each subdomain.
The data interaction between IFE and PIC meshes of different resolutions
is described in detail in \cite{RK_IEEE_2006_Plume_HybridGrid}.
Figures \ref{fig:2dmesh} and \ref{fig:3dmesh}
illustrate the 2-D and 3-D views of the domain decomposition
and different resolutions.

\subsection{Parallel Algorithm for IFE Field Solver}

For the parallel electrostatic field solver, Dirichlet-Dirichlet domain decomposition
with overlapping cells is used to distribute the subdomains
among multiple MPI processes \cite{Smith_DomainDecomp_1996}.
For each subdomain,
the IFE solver is the same as the sequential IFE method
with Dirichlet boundary conditions \cite{XHe_NMPDE2008_IFE,XHe_IFE_NonHomo_2011}.
These Dirichlet boundary conditions are imposed at the boundaries of the subdomain,
which are interior for the neighboring subdomains
(Figure \ref{fig:innerBC}, left).
Therefore, the field solution at respective neighboring subdomains are used as
Dirichlet boundary conditions for each subdomain.
Within each field-solve step, inner iterations are performed such that
the solutions of the overlapping cells are exchanged
and updated as the new Dirichlet boundary conditions
for the respective neighboring subdomains.
It is noted here that
since PIFE-PIC uses 3-D domain decomposition,
such MPI data exchange will be carried out
at guard cell nodes on ``surfaces''
(+/- neighbor in one direction, such as Rank 1 and Rank 2 in Figure \ref{fig:innerBC}),
``edges'' (+/- neighbor in two directions,
such as Rank 3 and Rank 6 in Figure \ref{fig:innerBC}),
and ``vertices'' (+/- neighbor in three directions).
We denote this level of iteration as the ``Domain Decomposition Method (DDM)'' iteration.
The relative error $e_\mathrm{rel}$ of DDM is defined with the $L^2$ norm as below:
\begin{equation}
e_\mathrm{rel} =
\frac{\norm{\phi_\mathrm{new} - \phi_{\mathrm{old}}}_{L^2}}{\|\phi_{\mathrm{old}}\|_{L^2}}
\end{equation}
where $\phi_\mathrm{new}$ and $\phi_\mathrm{old}$ denote solutions
at the new and old steps in the DDM iteration, respectively.

\subsection{Parallel Scheme for PIC Procedures}
\begin{figure}[ht!]
\centering
\includegraphics[width=0.80\textwidth]{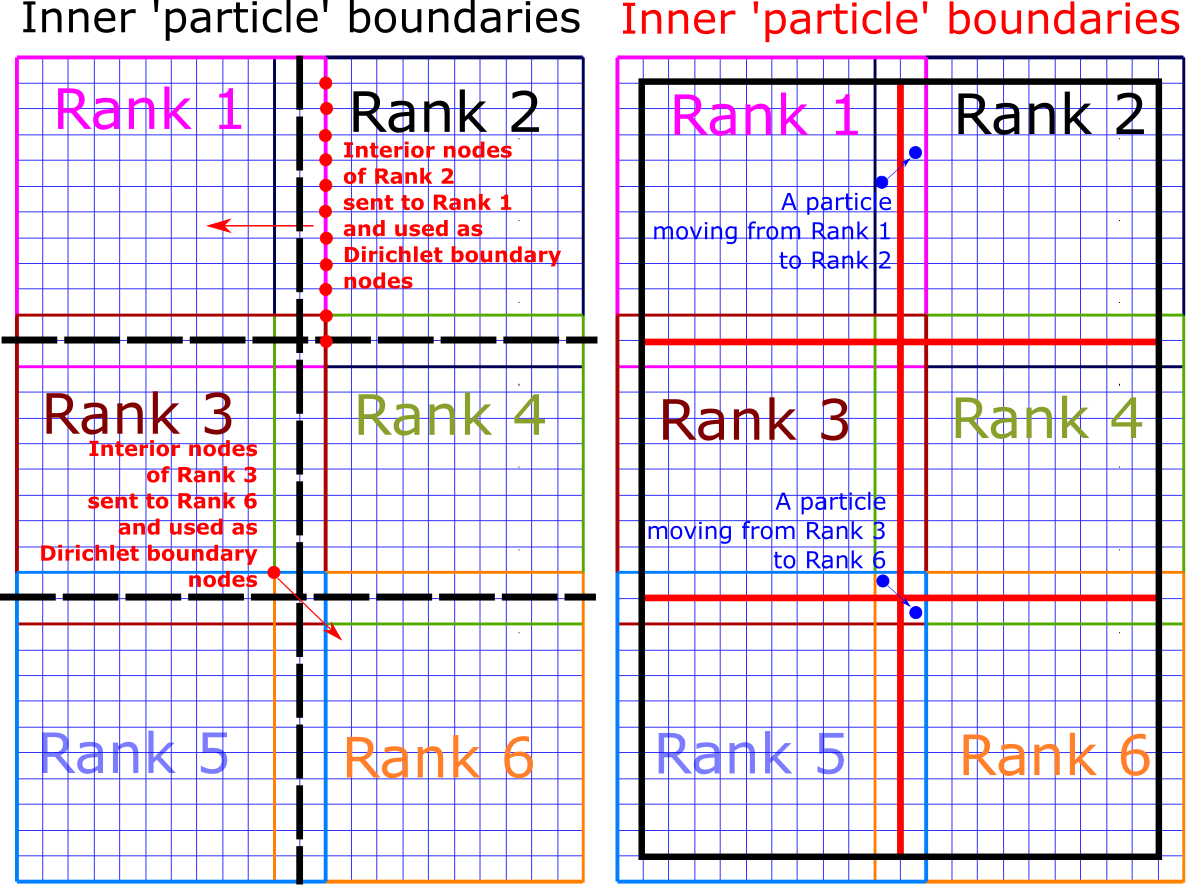}
\caption{MPI data exchange among neighboring subdomains within DDM iteration.
The thick edges (black and red) represent the boundaries of each subdomain without guard cells.
Left: for field-solve operations: at inner boundaries with guard cells,
the nodes at a certain subdomain's boundary (e.g., Rank 1's boundary nodes)
are also interior nodes of its neighboring subdomain (e.g., Rank 2).
Therefore, the field quantities stored on interior nodes of Rank 2
are sent to Rank 1 and used as Dirichlet boundary nodes.
Since PIFE-PIC has 3-D domain decomposition,
such MPI data exchange will be carried out
at guard cell nodes on ``surfaces''
(+/- neighbor in one direction, such as Rank 1 and Rank 2),
``edges'' (+/- neighbor in two directions, such as Rank 3 and Rank 6),
and ``vertices'' (+/- neighbor in three directions, not shown on this 2-D illustration).
Right: For particle-push operations:
each processor handles its own particles belonging to its domain
without guard cells (see Figure \ref{fig:pifepic:blocks}).
In particle-push, particles crossing the inner boundaries
are sent to the corresponding rank based on their destination positions.
Note such particle motion includes similar cases as data exchange for field-solve,
which are ``crossing one surface''
(+/- neighbor in one direction, such as Rank 1 and Rank 2),
``crossing an edge (two surfaces)''
(+/- neighbor in two directions, such as Rank 3 and Rank 6),
and ``crossing a vertex (three surfaces)''
(+/- neighbor in three directions, not shown on this 2-D illustration).
For charge-weighting, contributions from all neighboring subdomains are summed together
at respective inner boundary nodes.}
\label{fig:innerBC}
\end{figure}

In PIFE-PIC, simulation particles belonging to the same subdomain are stored together
on the processor that solves the electrostatic field of the same subdomain
(Figure \ref{fig:innerBC}, right).
In this sense, ``particle quantities'' and ``field quantities'' of each subdomain
are handled by the same processor.
Each processor (MPI rank) handles its own particles belonging to its domain
without guard cells (see Figure \ref{fig:pifepic:blocks}).
In particle-push, particles crossing the inner boundaries
are sent to the corresponding rank based on their destination positions.
Note that such particle motion includes similar cases as data exchange for field-solve,
which are ``crossing one surface''
(+/- neighbor in one direction, such as Rank 1 and Rank 2 in Figure \ref{fig:innerBC}),
``crossing an edge (two surfaces)''
(+/- neighbor in two directions, such as Rank 3 and Rank 6 in Figure \ref{fig:innerBC}),
and ``crossing a vertex (three surfaces)''
(+/- neighbor in three directions).

\subsection{Flowchart for PIFE-PIC}

\begin{figure}[ht!]
\centering
{\includegraphics[width=0.8\textwidth]{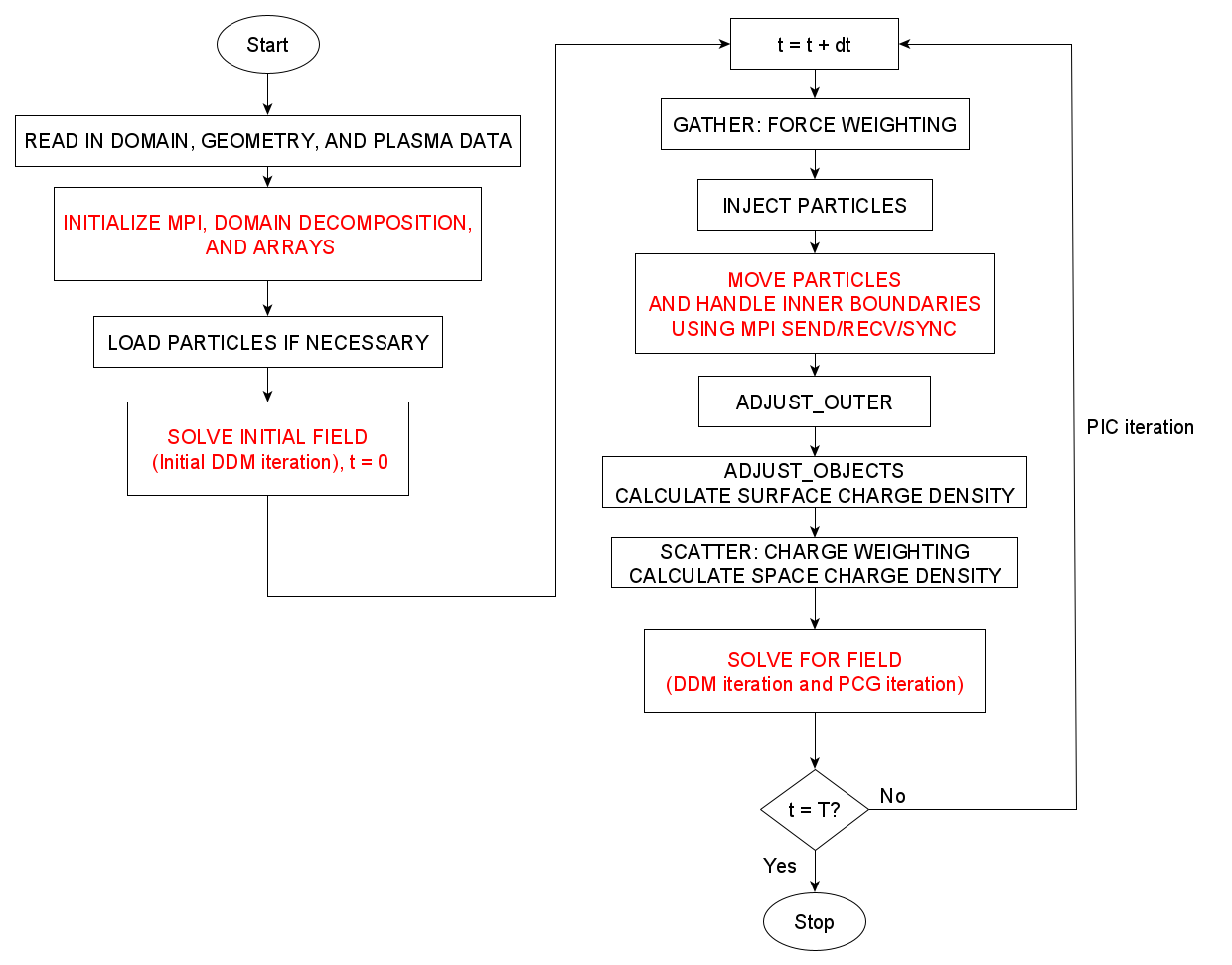}}
\caption{Flowchart of PIFE-PIC.}
\label{fig:flowchart:parallel}
\end{figure}

Figure \ref{fig:flowchart:parallel} shows the flowchart of PIFE-PIC.
The steps in red color are major steps involving MPI operations
associated with domain decomposition.
In total, there are three levels of iteration in PIFE-PIC.
The first level is the matrix-solving iteration
which uses the preconditioned conjugate gradient (PCG) algorithm (PCG level).
The second one checks the relative error in the iterations
of the domain decomposition method (DDM level).
The third one tracks the solution of each PIC step (PIC level).

\section{Code Validation}
\label{sec:validation}
\phantomsection

We apply the PIFE-PIC code to simulate the charging of a small dielectric
sphere immersed in a collisionless and stationary plasma
in the orbital-motion-limited (OML) sheath regime.
Successful validations of the serial IFE-PIC against analytic OML solutions are
presented in earlier work \cite{Han_ife_jcp2016,Han_ieee2016_charging}.

\subsection{Problem Description and Simulation Setup}

We consider a stationary, collisionless hydrogen plasma
of equal ion and electron temperatures ($T_i = T_e$).
The analytic expressions for ion and electron densities in the plasma are given by
the revised OML theory \cite{tang_pop2014_oml,delzanno_pop2015_oml_vs_pic}.
Therefore, the analytic potential profile near the sphere can be numerically solved
from Poisson's equation in spherical coordinates.

\subsubsection{Computation Domain and Mesh}

\begin{figure}[ht!]
\centering
\begin{subfigmatrix}{2} 
\subfigure[IFE mesh]%
{\label{fig:oml_mesh}
\includegraphics[width=0.45\textwidth]{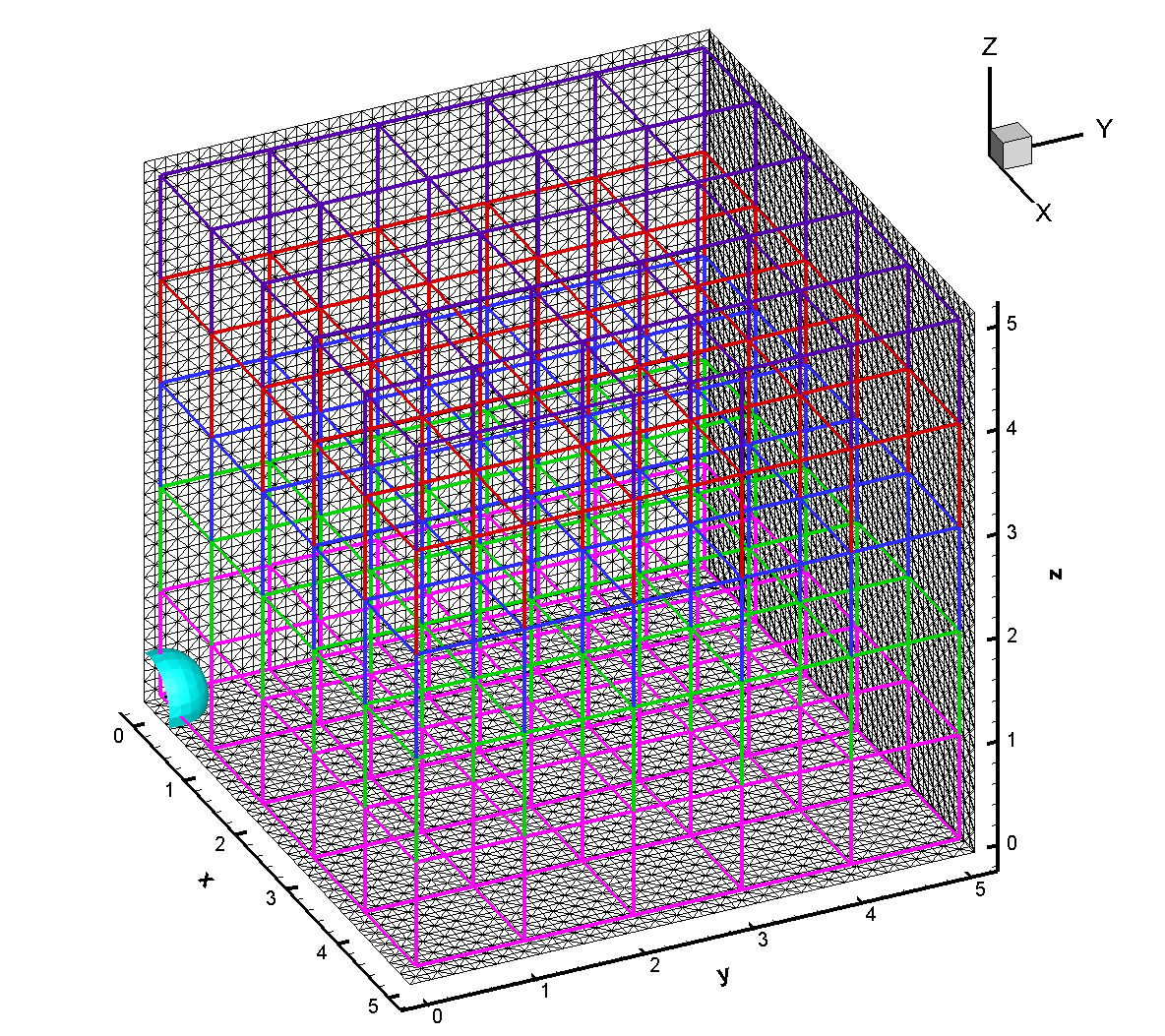}}
\subfigure[Simulation setup]%
{\label{fig:oml_setup}%
\includegraphics[width=0.45\textwidth]{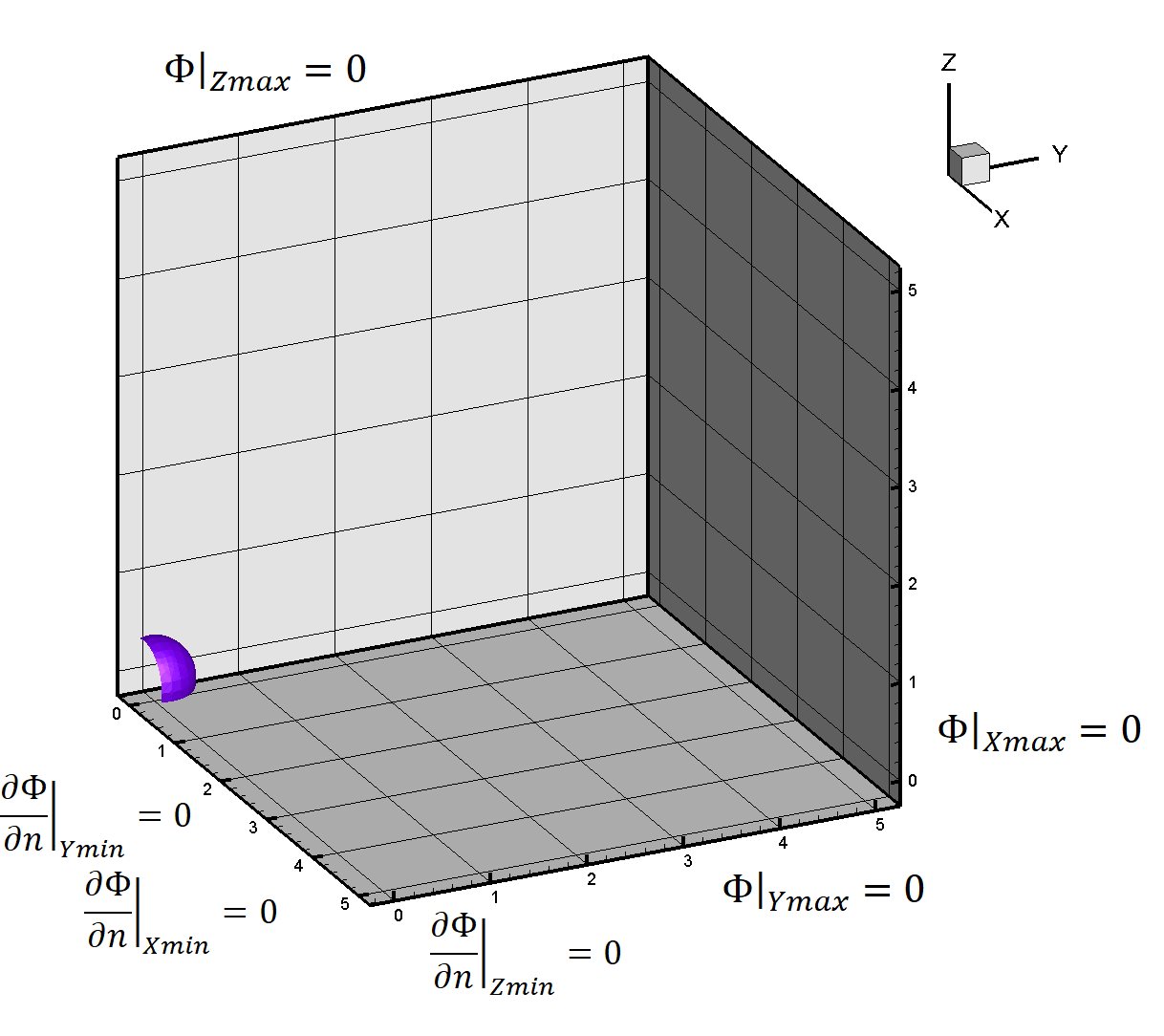}}
\end{subfigmatrix}
\caption{%
IFE mesh and setup used in the 3-D OML sheath problem for code validation.
In this example, the global domain is decomposed into
5$\times$5$\times$5 subdomains.
}
\label{fig:oml_mesh_setup}
\end{figure}

In the simulation, we use a computation domain of a 5\X 5\X 5
Debye cube with a globally uniform PIC mesh
with the size of $h = 0.1 \lambda_D$ in all dimensions,
where $\lambda_D$ is the Debye length of the plasma.
The entire simulation domain has 50\X 50\X 50 = 125,000 PIC cells
which is 125,000\X 5 = 625,000 tetrahedral FE/IFE cells
as each cuboid PIC cell is partitioned into 5 tetrahedral FE/IFE cells
in 3-D IFE-PIC \cite{Han_ife_jcp2016,Han_ieee2016_charging}.
The IFE mesh size is also globally uniform and the same as that of the PIC mesh.
The dielectric sphere is centered at $(0,0,0)$ with a radius of $R_s = 0.401$.
Due to symmetry in all three dimensions, only $1/8$ of the sphere is included in the domain.
The entire domain is partitioned into 5$\times$5$\times$5 subdomains
with each subdomain computed by one MPI process.
Figure \ref{fig:oml_mesh_setup} shows the 3-D IFE mesh
and setup used in the simulation.

\subsubsection{Field Setup}
At $X_{\max}$, $Y_{\max}$, and $Z_{\max}$ boundaries, the potentials are set to $0$
as the reference potential.
At $X_{\min}$, $Y_{\min}$, and $Z_{\min}$ boundaries, zero-Neumann boundary conditions
are applied due to symmetry
(Figure \ref{fig:oml_setup}).
The relative permittivity of the sphere is set to $4$.
The floating potential of the sphere
is calculated from the non-homogeneous flux jump condition at the sphere surface.

\subsubsection{Particle Setup}
The simulation is carried out using the realistic ion-to-electron mass ratio of
$m_i / m_e = 1836$.
Particles are pre-loaded into the domain before the initial field solution,
and injected into the domain at
$X_{\max}$, $Y_{\max}$, and $Z_{\max}$ within each PIC step.
Particles hitting the $X_{\min}$, $Y_{\min}$, and $Z_{\min}$ boundaries are reflected
due to symmetry.
Particles hitting the $X_{\max}$, $Y_{\max}$, and $Z_{\max}$ are absorbed and removed
from the simulation.
The normalized time step size was set to be 0.01.
There were 125 particles (5\X 5\X 5) per species,
per cell being loaded/injected into the domain.

\subsection{Simulation Results}

The simulation of the validation case finished in about 2 hours
for a total of 50,000 PIC steps
on the \emph{Foundry} cluster provided
by the Center of High-Performance Computing Research
at Missouri University of Science and Technology.
The computing nodes are configured with
Dell C6525 nodes each having
four node chassis with each node containing dual 32-core AMD EPYC Rome 7452 CPUs
with 256 GB DDR4 RAM and six 480GB SSD drives in RAID 0.
All other simulations presented in this work
were also carried out on the same cluster.


For this test case, the maximum number of PCG iterations was set to
60  with a tolerance \coloring{(for relative residual)} of \num{1e-6},
the max number of initial DDM iterations
(solving the initial electrostatic field before main PIC loop starts)
was set to 150
and the max number of DDM iterations at each PIC iteration step was set to
50 with a tolerance of \num{1e-2}.
The simulation was set to run 50,000 PIC steps.

\subsubsection{Initial Field Solution}
The initial field solution (the zeroth PIC step)
took about 100 DDM iterations which are more than
what is needed at each step of the main PIC loop,
to converge in terms of the relative error \num{1e-2}.
The idea of setting a relatively larger DDM iteration number
is to obtain a better initial field for the main PIC loop.
Since the initial field was solved only once,
the extra DDM iterations contributed little to the overall wall-clock time
of the entire simulation.

\subsubsection{Solution History of Main PIC Loop}

\begin{figure}[ht!]
\centering
\includegraphics[width=0.85\textwidth]%
{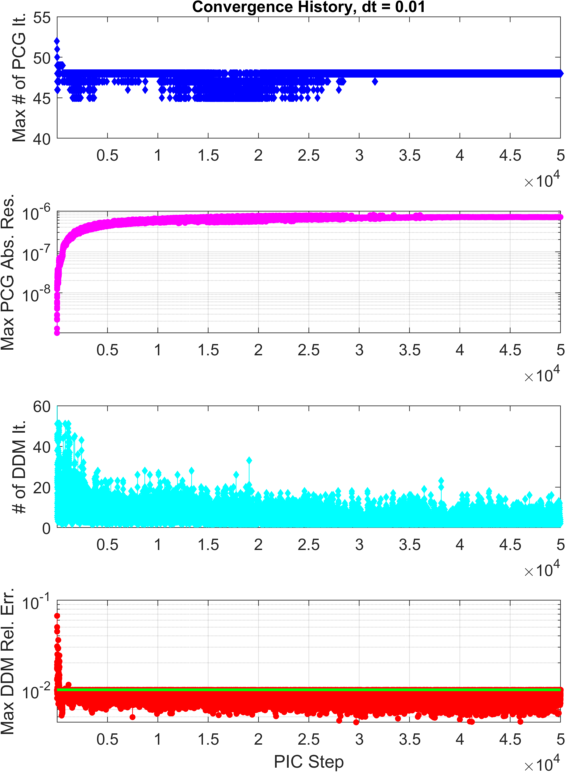}
\caption{%
Field convergence history of the code validation test case,
PCG absolute residual and DDM relative error.
The green line on max DDM relative error plot is the DDM tolerance.
}
\label{fig:convergence_history}
\end{figure}

Figure \ref{fig:convergence_history} shows the field solution
convergence history
including the \coloring{max absolute PCG residual and max DDM relative error}
as a function of PIC steps in the main PIC loop.
A few phenomena are observed here:
\begin{enumerate}
\item For most PIC steps, PCG took about 45-50 iterations
to reach the tolerance of \num{1e-6}.
The ``max'' in the vertical axis stands for ``max among all subdomains''
(first plot);

\item The maximum PCG absolute residual of the matrix solver has been maintained
below \num{1e-6} (second plot);

\item At early PIC steps, most DDM steps took about tens of iterations to converge
below \num{1e-2},
while later on as PIC approaches steady state,
most DDM steps were able to converge within about 10$\sim$15 iterations
(third and forth plots).

\end{enumerate}

\begin{figure}[ht!]
\centering
\includegraphics[width=0.80\textwidth]%
{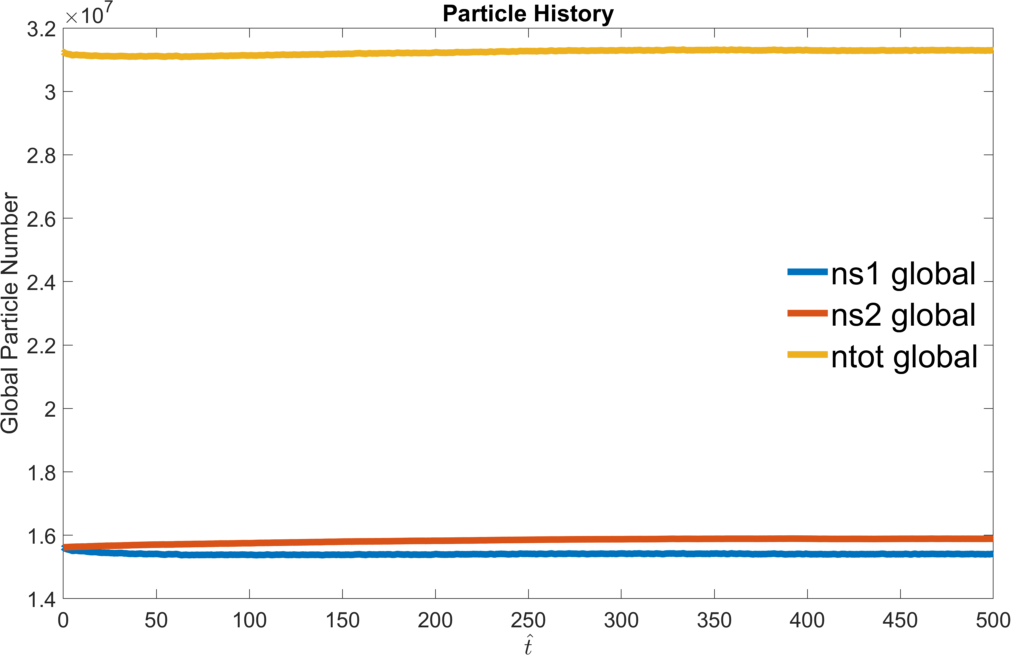}
\caption{%
Global particle history of the code validation test case.
``ns1'' represents the number of Species \#1 particles which is the number of electrons,
while ``ns2'' represents the number of Species \#2 particles
which is the number of ions.
``ntot'' is the total number of particles (electrons plus ions).
}
\label{fig:particle_history}
\end{figure}

Figure \ref{fig:particle_history} shows the global particle number history.
At the steady state,
there are approximately \num{1.56e7} particles in the entire global domain.
It is also shown that the numbers of particles
reached steady state at normalized simulation time of about $\hat{t} = 125$.

\subsubsection{Comparison with Analytic Solution}

\begin{figure}[ht]
\centering
\begin{subfigmatrix}{2} 
\subfigure[Potential profile]%
{\includegraphics[width=0.45\textwidth]{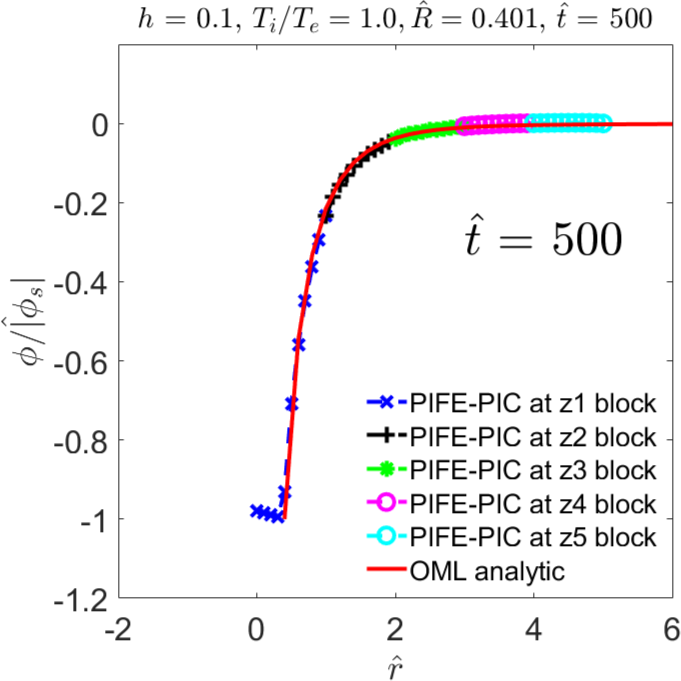}}
\subfigure[3-D potential contour]%
{\includegraphics[width=0.45\textwidth]{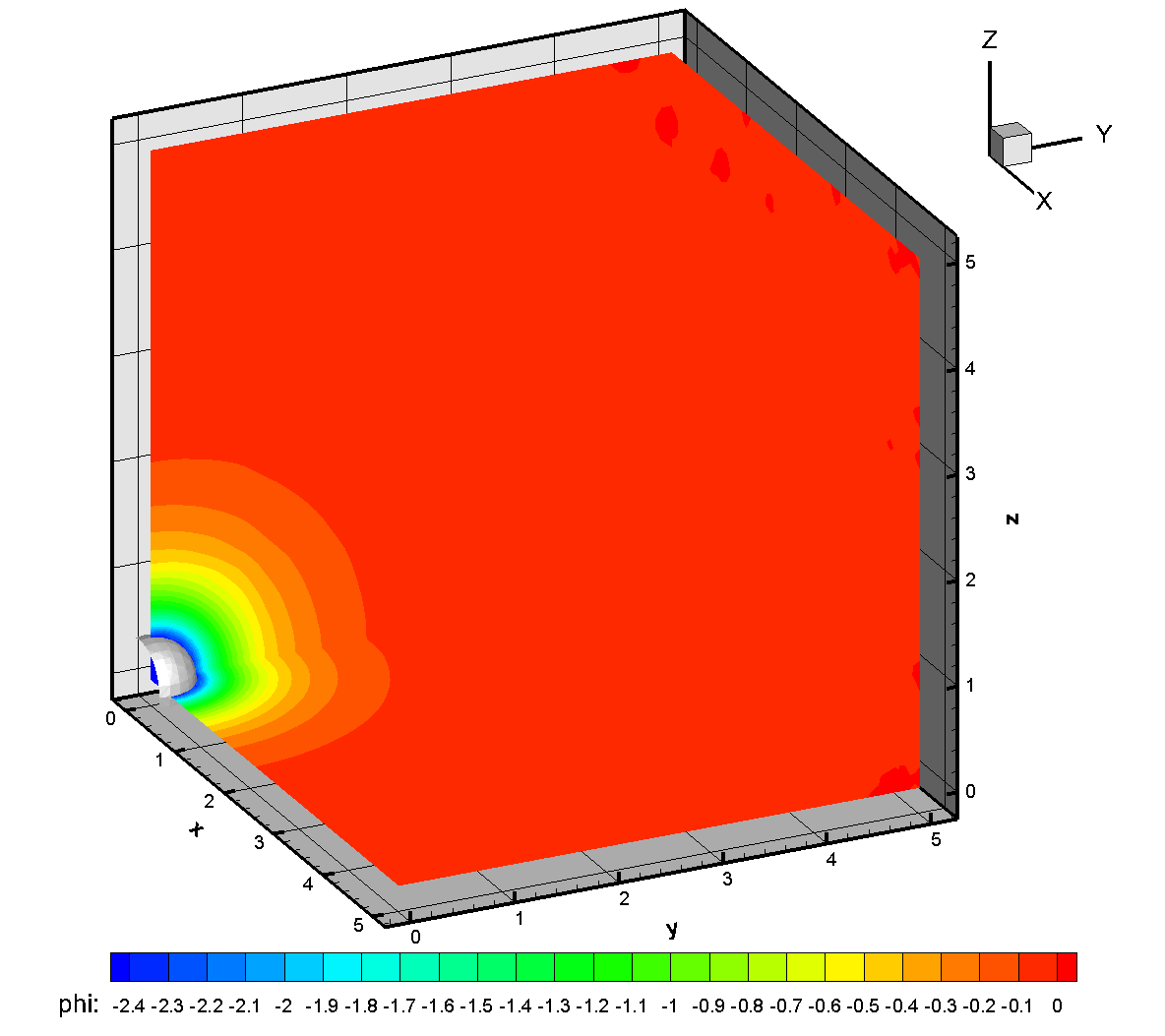}}
\end{subfigmatrix}
\caption{%
Validation of OML sheath solution: PIFE-PIC vs. analytic and 3-D potential contours.}
\label{fig:results}
\end{figure}

Figure \ref{fig:results} shows the comparison between PIFE-PIC simulation results
against analytic solution for the OML sheath problem as well as a 3-D potential contour.
The potential profile agrees very well with the analytic solution,
as also shown in earlier work with the serial IFE-PIC
\cite{Han_ieee2016_charging,Han_ife_jcp2016}.

\subsection{Performance Profiling}

Table \ref{tab:profiling} shows the detailed timer profile of PIFE-PIC
on the validation simulation
for the entire 50,000 PIC steps:
total wall-clock time and
percentage of total wall-clock time
for each main component of the PIFE-PIC procedures, namely,
``gather'',
``particle-push'',
``particle-push-comm'' (adjust particles at local boundaries and communication among subdomains),
``scatter'',
``field-solve'',
``field-solve-phibc'' (communication among subdomains
and update local potential boundary conditions),
and ``other'' (including particle injection at global boundaries
and particle collection at objects).
The percentage breakdown shows
the procedures of `particle-push' and `field-solve'
took the majority of the computing (wall-clock) time.
The computing time of `particle-push' essentially depends on the number of simulation particles
in the domain,
which also affects the accuracy and smoothness of the source term for Poisson's equation.
Therefore, in practical PIC simulations,
large numbers of particles are preferred when computing resources are available.
The computing time of `field-solve' essentially
depends on 1) the size of each subdomain (number of mesh cells and nodes)
and 2) the number of DDM iterations.
The size of each subdomain can be determined by the domain decomposition configurations,
while the number of DDM iterations is affected by
the DDM relative error tolerance and the max number of DDM iterations.
In the following section, we vary the size of each subdomain
and max number of DDM iterations
to investigate the strong scaling performance of PIFE-PIC.

\begin{table}[ht!]
  \begin{center}
    \caption{Time percentage breakdown for all 50,000 PIC steps.}
    \vspace{-0.5cm}
    \label{tab:profiling}
\resizebox{\textwidth}{!}{%
    \begin{tabular}{c|c|c}
    \hline
    \hline
      \textbf{Computing step} & \textbf{Wall-clock time (s)}
      & \textbf{Percent of total wall-clock time (\%)}\\ %
    \hline
    \hline
      {Total wall-clock time } & {6672.31} & {100.00}\\ %
      {Total gather time} & 744.88 & 11.16 \\ 
      {Total particle-push time} & 3153.67 & 47.26 \\ 
 	{Total particle-push-comm (AdjustOuter local) time$^*$} & 1652.67 & 24.77$^*$ \\ 
      {Total scatter time} & 48.48 & 3.72 \\ 
      {Total field-solve time} & 2129.57 & 31.92 \\ 
	{Total field-solve-phibc (Update Phi BC) time$^{**}$} & 310.79 & 4.66$^{**}$ \\ 
      {Total other time} & 95.71 & 5.93 \\ 
      \hline
    \end{tabular}
}
  \end{center}
\footnotesize{$^*$ Included in the `particle-push time'}\\
\footnotesize{$^{**}$ Included in the `field-solve time'}
\end{table}

%
\section{Parallel Efficiency: Strong Scaling}
\label{sec:scaling}
\phantomsection

For most large-scale problems of practical interests,
the problem size is usually determined by the physical phenomena to be resolved.
Therefore, to test the parallel efficiency of PIFE-PIC,
we use the \emph{strong} scaling approach such that
the problem size is \emph{fixed} while the number of processors increases.
For this set of tests, the problem size was fixed
as a 10$\times$10$\times$10 Debye cube
with a globally uniform PIC mesh size of $h = 0.1 \lambda_D$ in all dimensions.
The entire simulation domain has 100$\times$100$\times$100 = 1 million PIC cells
(5 million tetrahedral FE/IFE cells)
{and about 54 million particles}.
For these runs,
the max number of PCG iteration was set to 150 with a tolerance of \num{1e-6}.
For the initial field solution,
the max number of DDM iteration was set to 100,
while for each step within the main PIC loop,
the max number of DDM iteration
was set to be 10 and 6 for two different groups
with same tolerance of \num{1e-2}.
The normalized time step size was set to be 0.01 and
all simulations ran for 20,000 PIC steps.
The speedup is defined as $S = {T_s}/{T_p}$,
where $T_s$ is the serial runtime and $T_p$ is the
parallel runtime on $p$ MPI processes.
The strong scaling parallel efficiency is then
defined as $E = S/p = {T_s}/{(p \cdot T_p)}$.
We chose two groups of configurations to test the parallel efficiency:

\begin{itemize}
\item Group I: Using at most 10 DDM iterations per main-loop PIC step;

\item Group II: Using at most 6 DDM iterations per main-loop PIC step.

\end{itemize}
The timer data was taken over all 20,000 PIC steps.
Table \ref{tab:scaling:config} lists the domain decomposition configurations
for each test case.
Table \ref{tab:scaling:results} lists the total wall-clock time,
speedup, and parallel efficiency of each case for both Group I and Group II.

\begin{table}[ht!]
\centering
\caption{Domain Decomposition Configurations for Strong Scaling Test Cases}
\label{tab:scaling:config}
\begin{tabular}%
{%
>{\centering}m{1.5cm}
|>{\centering}m{2.8cm}
|>{\centering}m{2.8cm}
|>{\centering\arraybackslash}m{2.8cm}
}%
\hline
\# of subdomains &DD Configurations
&Size of smallest subdomain (cells)
&Size of biggest subdomain (cells)\\
\hline
1 (serial) &1$\times$1$\times$1 &100$\times$100$\times$100 &100$\times$100$\times$100 \\
%
64 &4$\times$4$\times$4 &25$\times$25$\times$25 &25$\times$25$\times$25 \\
%
80 &4$\times$4$\times$5 &20$\times$20$\times$20 &25$\times$25$\times$25 \\
%
100 &4$\times$5$\times$5 &20$\times$20$\times$20 &25$\times$25$\times$25 \\
%
125 &5$\times$5$\times$5 &20$\times$20$\times$20 &20$\times$20$\times$20 \\
%
%
180 &5$\times$6$\times$6 &15$\times$15$\times$15 &20$\times$20$\times$20 \\
%
216 &6$\times$6$\times$6 &15$\times$15$\times$15 &17$\times$17$\times$17 \\
\hline
\end{tabular}
\end{table}

\begin{table}[ht!]
\centering
\caption{Strong Scaling Test Results}
\label{tab:scaling:results}
\begin{tabular}%
{%
>{\centering}m{1.5cm}
||>{\centering}m{2.0cm}
|>{\centering}m{1.6cm}
|>{\centering}m{1.5cm}
||>{\centering}m{2.0cm}
|>{\centering}m{1.6cm}
|>{\centering\arraybackslash}m{1.5cm}
}%
\hline
\# of subdomains
&Total time $T_{\mathrm{I}}$ (min)
&Speedup $S_{\mathrm{I}}$
&Efficiency $E_{\mathrm{I}}$ (\%)
&Total time $T_{\mathrm{II}}$ (min)
&Speedup $S_{\mathrm{II}}$
&Efficiency $E_{\mathrm{II}}$ (\%)\\
\hline
1 (serial) &12,509 &1 &100 &12,509 &1 &100\\
64 &196 &63.78 &99.66 &177 &70.54 &110.22\\
80 &160 &78.42 &98.03 &141 &88.81 &111.01\\
100 &149 &83.89 &83.89 &117 &106.87 &106.87\\
125 &142 &88.26 &70.61 &109 &114.76 &91.81\\
180 &102 &123.24 &68.47 &77 &162.93 &90.51\\
216 &93 &134.01 &62.04 &70 &178.16 &82.48\\
\hline
\end{tabular}
\end{table}

\begin{figure}[ht]
\centering
\begin{subfigmatrix}{2}
\subfigure[Group I]%
{\includegraphics[width=0.75\textwidth]{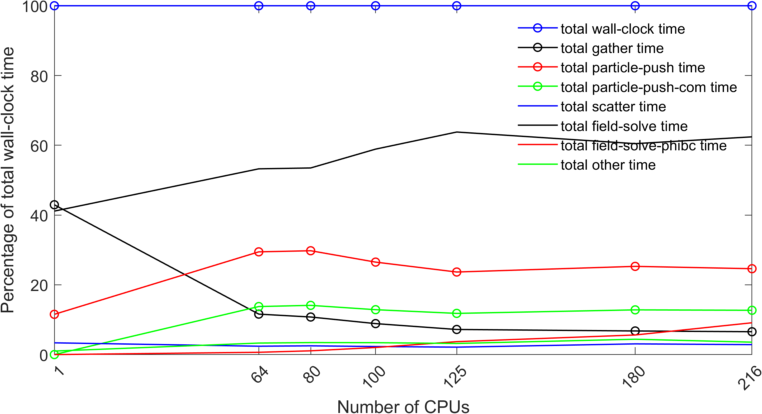}}
\subfigure[Group II]%
{\includegraphics[width=0.75\textwidth]{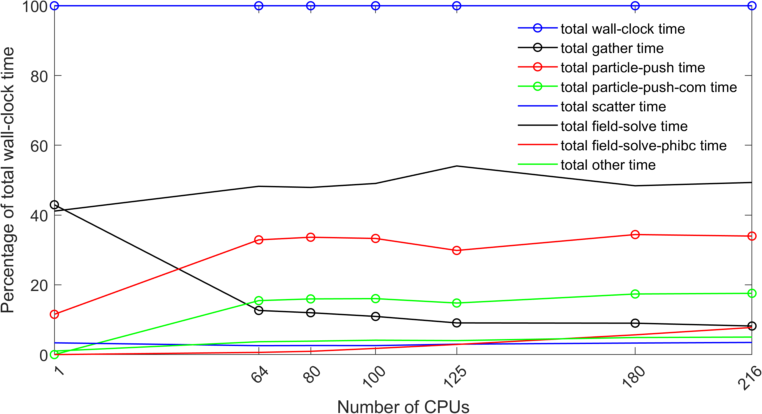}}
\end{subfigmatrix}
\caption{%
Percentage of total wall-clock time
for all PIC steps.
}
\label{fig:scaling:percentage}
\end{figure}

Figure \ref{fig:scaling:percentage}
plot the percentage of total wall-clock time 
to show the performance of PIFE-PIC
across different number of processors for Group I and Group II.
The percentage of total wall-clock time breakdown
is fairly consistent across all parallel configurations.
It also shows the majority of the wall-clock time
was always consumed by the field-solve step.

%

%

\section{Application to Lunar Crater Charging}
\label{sec:lunar}
\phantomsection

In this section, we apply PIFE-PIC to simulate the plasma charging
at a lunar crater under average solar wind (S.W.) conditions
to demonstrate the large-scale simulation capability of PIFE-PIC.
In the following, we will first briefly describe the lunar surface charging problem,
then introduce the setup of the simulation,
and finally present the results and discussion.

\subsection{Problem Description}

The problem considered is solar wind plasma charging near the lunar surface,
specifically, near the lunar craters at the terminator region
for lunar exploration missions.
The Moon is directly exposed
to the solar radiation and various space plasma environments
which directly interact with the lunar surface.
A direct consequence of such interactions is surface charging.
Observations have found that the potential of the sunlit surface is typically
a few tens of volts positive with respect to ambient due to photoelectron emission,
while that of the surface in shadow can be hundreds to thousands of volts negative
because of the hot electron flux from ambient plasma
that can dominate the charging process
\cite{Apollo17_PSR_1973,
Freeman_moon1975_electric,
Halekas_grl2007_lunar_charging,
Halekas_jgr2008_lp,
Halekas_pss2011_new_views,Reasoner_jgr1972_lunar_photoe_layer,Stubbs_LunarSurfaceCharging_2005,
Willis_1973_phe_se}.
Both solar illumination and plasma flow
can have a substantial influence on lunar surface charging.
At the lunar terminator,
the rugged surface terrain, such as that near a crater,
generates localized plasma wakes and shadow regions which
can lead to strong differential charging at the surface
\cite{Berg_epsl1978_lunar_terminator_config,Poppe_icarus2012_topography,
JW_IEEE_2008_LunarDust}.
Both the localized plasma flow field and the charged lunar surface are expected
to have substantial influence on the charging of spacecraft/landers/rovers/habitats
for future surface missions.

The lunar surface is covered by the lunar regolith layer which separates the solid bedrock
from the plasma environment.
The regolith layer in most areas is about 4 to 20 meters thick
\cite{lunarsourcebook_ch7_regolith,shkuratov_icarus2001_regolith}.
A complete model of plasma charging on the lunar surface
needs to explicitly take into account
the properties of the regolith layer, such as permittivity, layer thickness,
and the lunar electrical ground.

The serial version of IFE-PIC method has been successfully applied to
simulations of lunar plasma charging
\cite{Han_jsr_2018_lunar}.
In order to illustrate the high performance computing capability
of the PIFE-PIC package in this paper,
we 
apply PIFE-PIC to a much larger scale parallel simulation with
a larger simulation domain including a lunar crater
and much more simulation particles.
%
%
The plasma environment is chosen to be the average solar wind
and photoelectron parameters at the lunar surface \cite{JW_IEEE_2008_LunarDust},
as shown in Table \ref{tab:solarwind}.
It is noted here that the Debye length of photoelectrons at 90$^\circ$ Sun elevation angle
(1.38 m) is used as the reference length to normalize spatial dimensions in PIFE-PIC.

\begin{table}[ht!]
\begin{center}
\caption{Average solar wind and photoelectron
(at 90$^\circ$ Sun elevation angle) parameters}
\label{tab:solarwind}
\begin{tabular}{m{0.18\textwidth}
               |>{\centering}m{0.10\textwidth}|>{\centering}m{0.11\textwidth}
               |>{\centering}m{0.11\textwidth}|>{\centering}m{0.16\textwidth}
               |>{\centering\arraybackslash}m{0.1\textwidth}}
    \hline
    \hline
    & {Number density \textit{n}, cm$^{-3}$}
    & {Drifting velocity \textit{v$_d$}, km/s}
    & {Thermal velocity \textit{v$_t$}, km/s}
    & {Temperature \textit{T}, eV}
    & {Debye length \textit{$\lambda_D$}, m}\\ %
    \hline
    {S.W. electron} & 8.7 & 468 & 1453 & 12 & 8.73\\ 
    {S.W. ion} & 8.7 & 468 & 31 & 10 & N/A$^*$\\ 
    {Photoelectron} & 64 & N/A$^*$ & 622 & 2.2 & 1.38\\ 
    \hline
\end{tabular}

\end{center}
\footnotesize{$^*$ N/A denotes ``not applicable''}
\end{table}

\subsection{Simulation Setup}

\subsubsection{Lunar Crater Geometry and Simulation Domain}
\begin{figure}[ht!]
\centering
\begin{subfigmatrix}{2} 
\subfigure[The geometry of the lunar crater realized in PIFE-PIC.
Color contours show the ``sunlight index'' indicating the inner product
of Sun vector (10$^{\circ}$ above the ground in the $X-Z$ plane)
and local surface normal vector.]%
{\label{fig:lunarcrater}\includegraphics[width=0.45\textwidth]%
{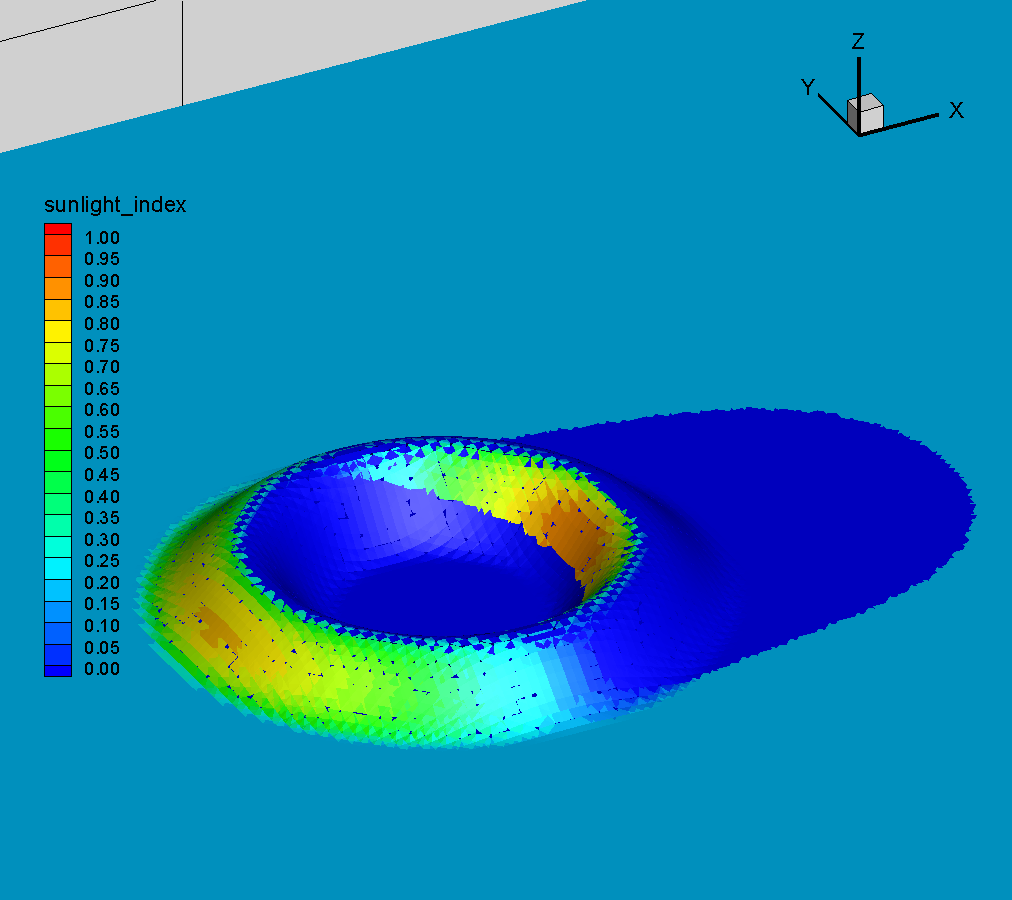}}
\subfigure[The simulation domain including the lunar bedrock (below the blue layer) and
the lunar regolith (between the blue and green layers).
The light-blue edges show the domain decomposition
(8\X 4\X 4 = 128 MPI processes).]%
{\label{fig:lunardomain}\includegraphics[width=0.45\textwidth]%
{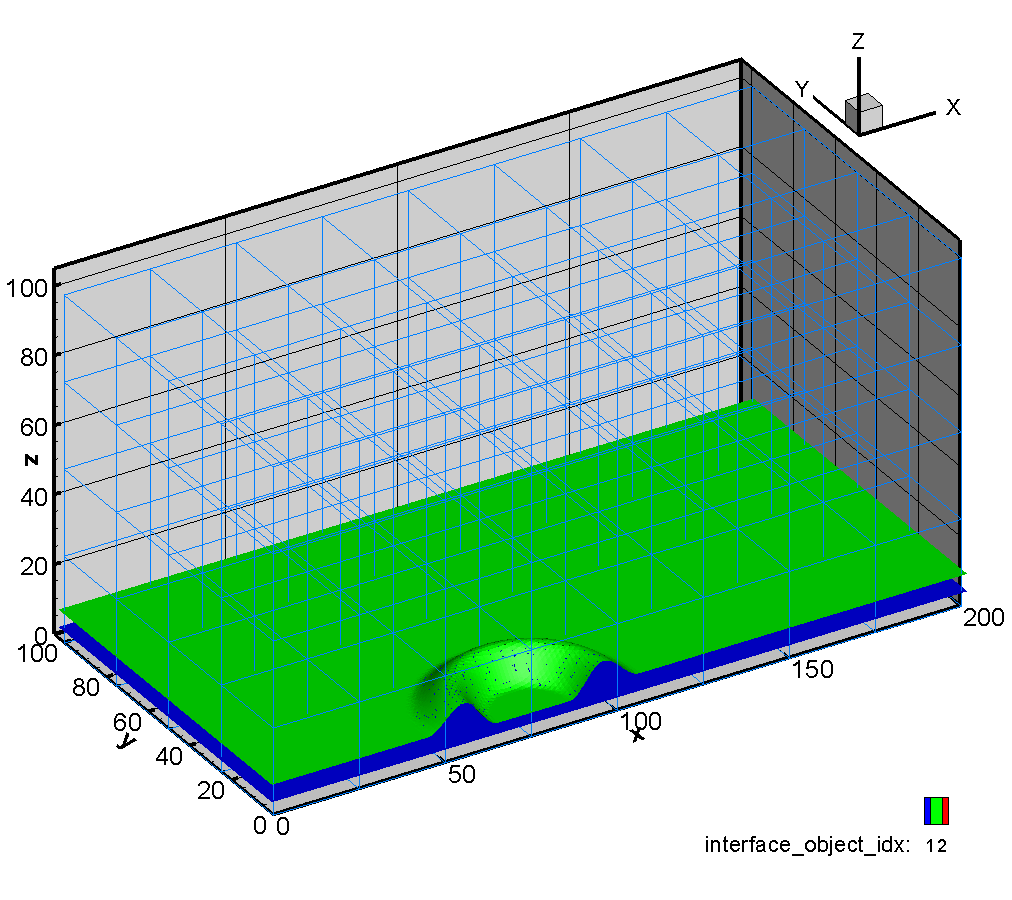}}
\end{subfigmatrix}
\caption{The lunar crater geometry and simulation domain.}
\label{fig:lunar:domain}
\end{figure}

In PIFE-PIC, the geometry of the lunar crater is realized through an algebraic equation
describing the surface terrain in the form of $z = z(x,y)$
where $z$ denotes the surface height.
For the lunar crater considered here,
the shape is realized by a few characteristic parameters
such as inner-rim radius, outer-rim radius, depth, rim height, etc.
(Figure \ref{fig:lunarcrater})
according to the \emph{Lunar Sourcebook} \cite{lunarsourcebook_1991}.
The specific diameter of a real lunar crater can be measured
through NASA Jet Propulsion Laboratory's website, Moon Trek \cite{moontrek}.
The crater considered in this study has these characteristic dimensions:
inner-rim radius 10.5$\times$1.38 = 14.49 m,
top-rim radius 20.2$\times$1.38 = 27.88 m,
outer-rim radius 30.9$\times$1.38 = 42.64 m,
and top height 6.7$\times$1.38 = 9.25 m.
Details of the approach to set up the lunar crater geometry is given in
Ref. \cite{Lund_AIAA_2020_1549_LunarCrater}.

The simulation domain has 200$\times$100$\times$100 = 2 million PIC cells
(10 million tetrahedral FE/IFE cells)
including {half} of the lunar crater due to symmetry with respect to
the $X$-$Z$ plane at $y=0$
(Figure \ref{fig:lunarcrater}).
Each PIC cell is a 1.38$\times$1.38$\times$1.38 cube.
In physical units, the domain size is approximately
276 m by 138 m by 138 m.
At the $Z_{\min}$ boundary,
the simulation domain includes a layer of
the lunar bedrock with a thickness of
$L_{\mathrm{bedrock}}$ = 4.5$\times$1.38 = 6.21 m.
On top of the bedrock is a layer of dielectric regolith with a thickness
of $L_{\mathrm{regolith}}$ = (9.5 - 4.5)$\times$1.38 = 6.9 m.
The relative permittivities of the lunar regolith layer and the bedrock
are taken to be
$\epsilon_{\mathrm{regolith}} = 4$
and $\epsilon_{\mathrm{bedrock}} = 10$,
respectively \cite{lunarsourcebook_1991}.
3-D domain decomposition of 8\X 4\X 4 (total 128 MPI processes)
is used to run the simulation (Figure \ref{fig:lunardomain}).

\subsubsection{Particle and Field Boundary Conditions}

Particles representing solar wind ions and electrons
are pre-loaded and injected into the domain
with an angle of 10$^{\circ}$ towards the surface
in the $X$-$Z$ plane (Figure \ref{fig:lunarcrater}).
Particles representing photoelectrons are generated at the sunlit regions
according to the local sunlight index.
At the global $X_{\min}$, $X_{\max}$, $Y_{\max}$, and $Z_{\max}$ domain boundaries,
ambient solar wind particles are injected.
Particles hitting the global $Y_{\min}$ boundary are reflected due to symmetry.
Particles hitting the lunar surface are collected and their charges are accumulated
to calculate surface charging.

Dirichlet boundary condition of $\Phi = 0$ is applied
at the $Z_{\max}$ boundary (the unperturbed solar wind), whereas
Neumann boundary condition of zero electric field is applied on
all other five domain boundaries.
The PCG max iterations was set to 150 with a tolerance of \num{1e-6}
\coloring{(for relative residual)}.
The max number of DDM iteration for initial field solution was set to 800
and the max number of DDM iteration for each step
within the main PIC loop
was set to 200 with a tolerance of \num{1e-3}.
The simulation ran for 20,000 PIC steps.


\subsection{Convergence History}

\begin{figure}[ht!]
\centering
\begin{subfigmatrix}{1} 
\subfigure[Field solution PCG absolute residual and DDM relative error history.]%
{\label{fig:lunar:convergence:residual}
\includegraphics[trim={3cm 0 3cm 0}, clip, width=0.75\textwidth]%
{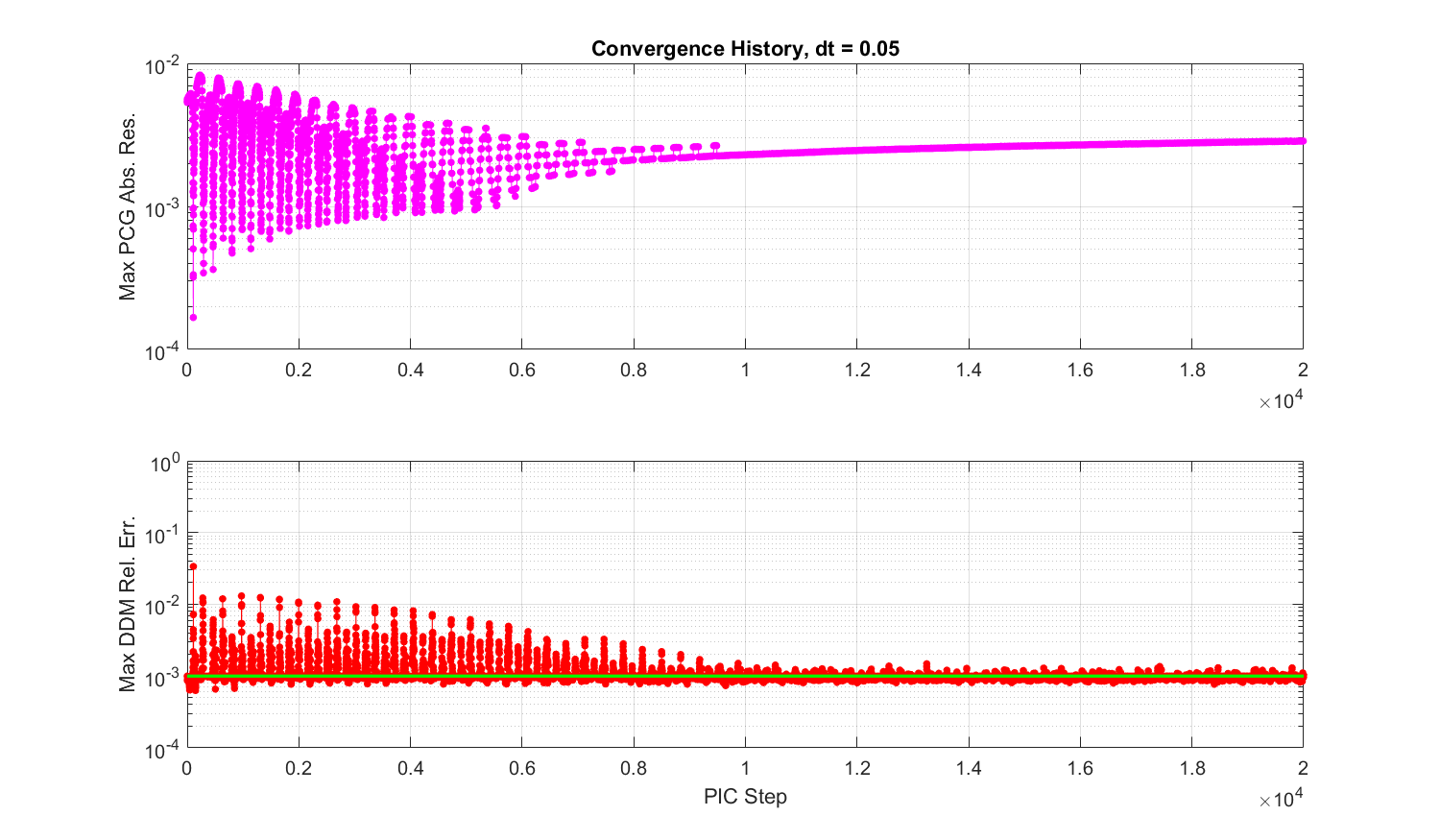}}
\subfigure[Global particle number history.]%
{\label{fig:lunar:convergence:particle}
\includegraphics[trim={3cm 0 3cm 0}, clip, width=0.75\textwidth]%
{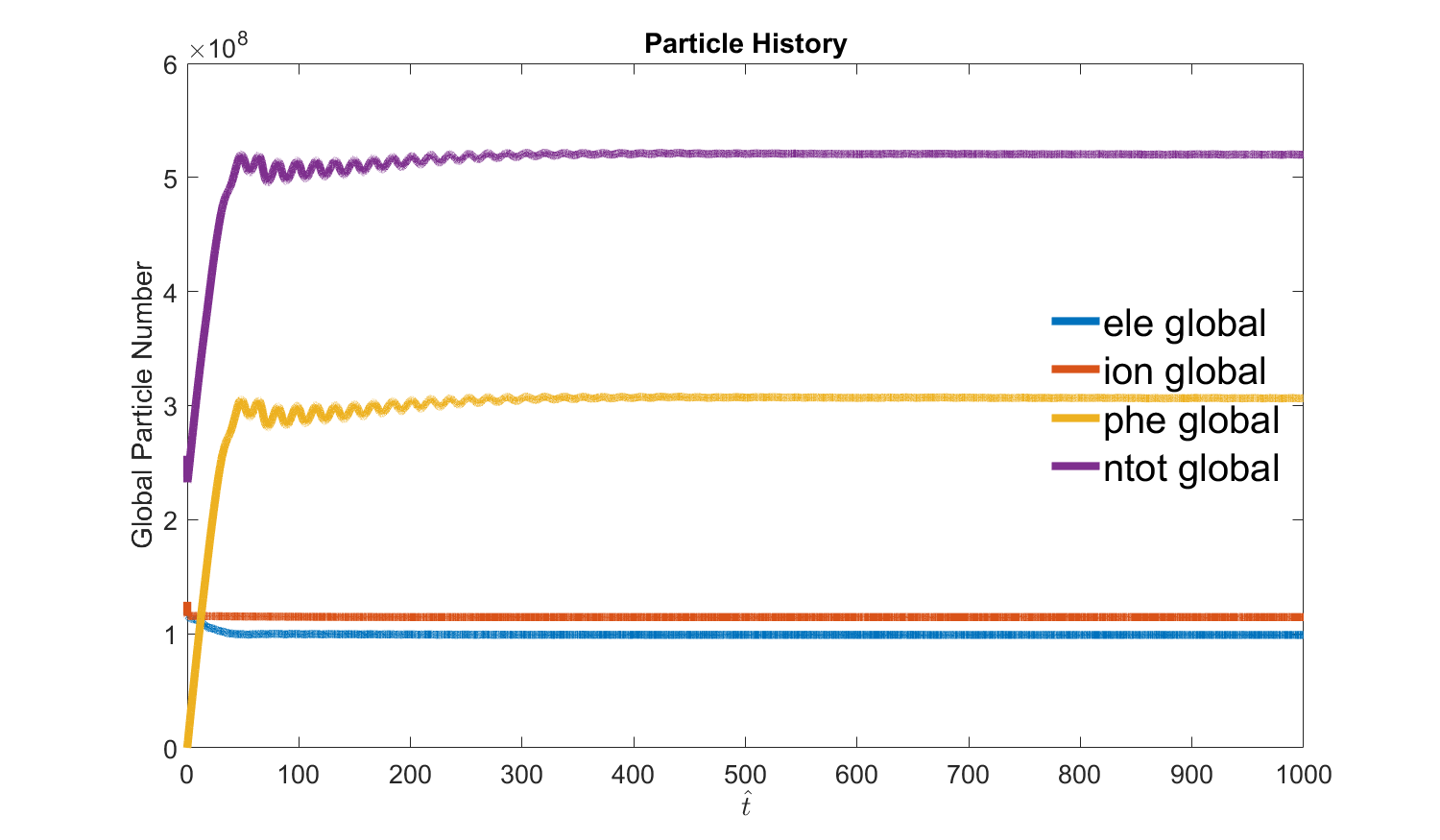}}
\end{subfigmatrix}
\caption{The lunar crater simulation convergence history.}
\label{fig:lunar:convergence}
\end{figure}

The run took about 109 hours to finish 20,000 PIC steps with the time step size of
0.05 (total simulation time till $\hat{t} = 1,000$).
Figure \ref{fig:lunar:convergence} shows the convergence history
of the lunar crater charging simulation
including the \coloring{max absolute PCG residual and max DDM relative error}
and particle number histories.
It is shown that the field solution residuals and relative errors started to level off
near PIC step of 10,000 ($\hat{t} = 500$),
and at steady state, the entire domain had about 520 million particles.
The results presented below are taken at $\hat{t} = 1,000$.

\subsection{Surface Charging Results}

\begin{figure}[ht!]
\centering
\begin{subfigmatrix}{2} 
\subfigure[Solar wind ion density contours.]%
{\includegraphics[width=0.45\textwidth]%
{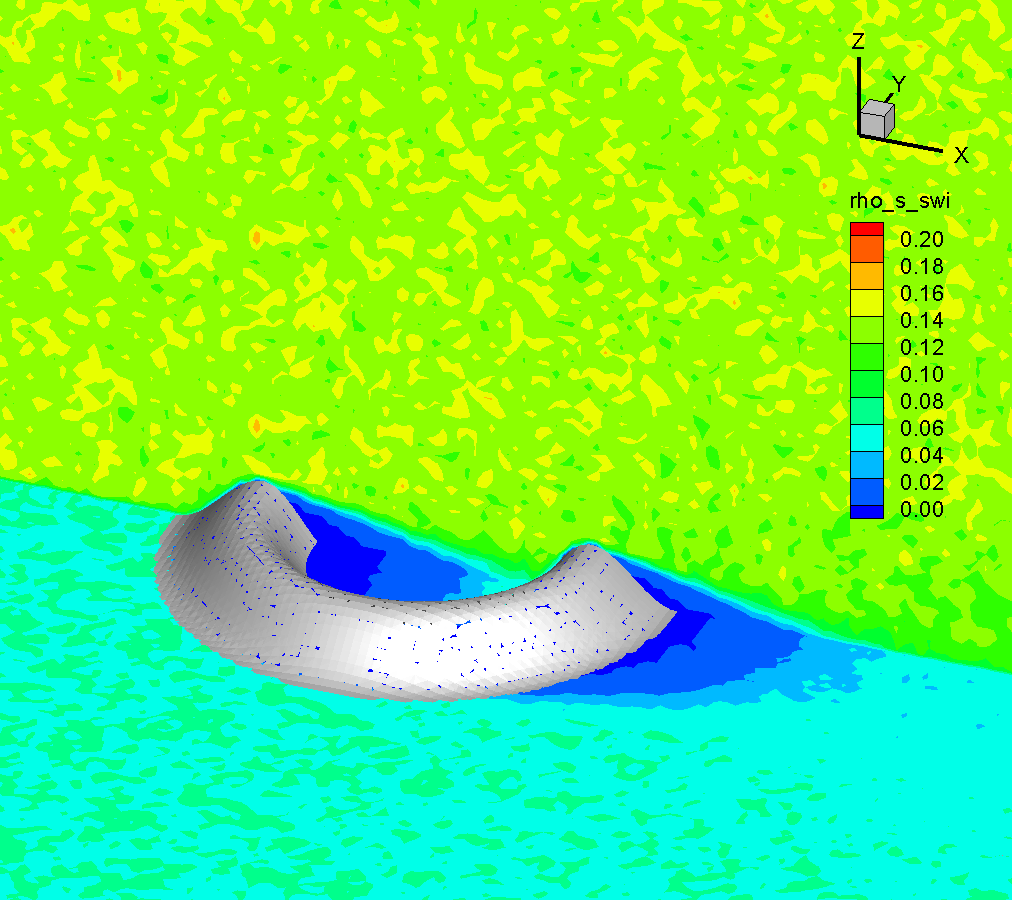}}
\subfigure[Solar wind electron density contours.]%
{\includegraphics[width=0.45\textwidth]%
{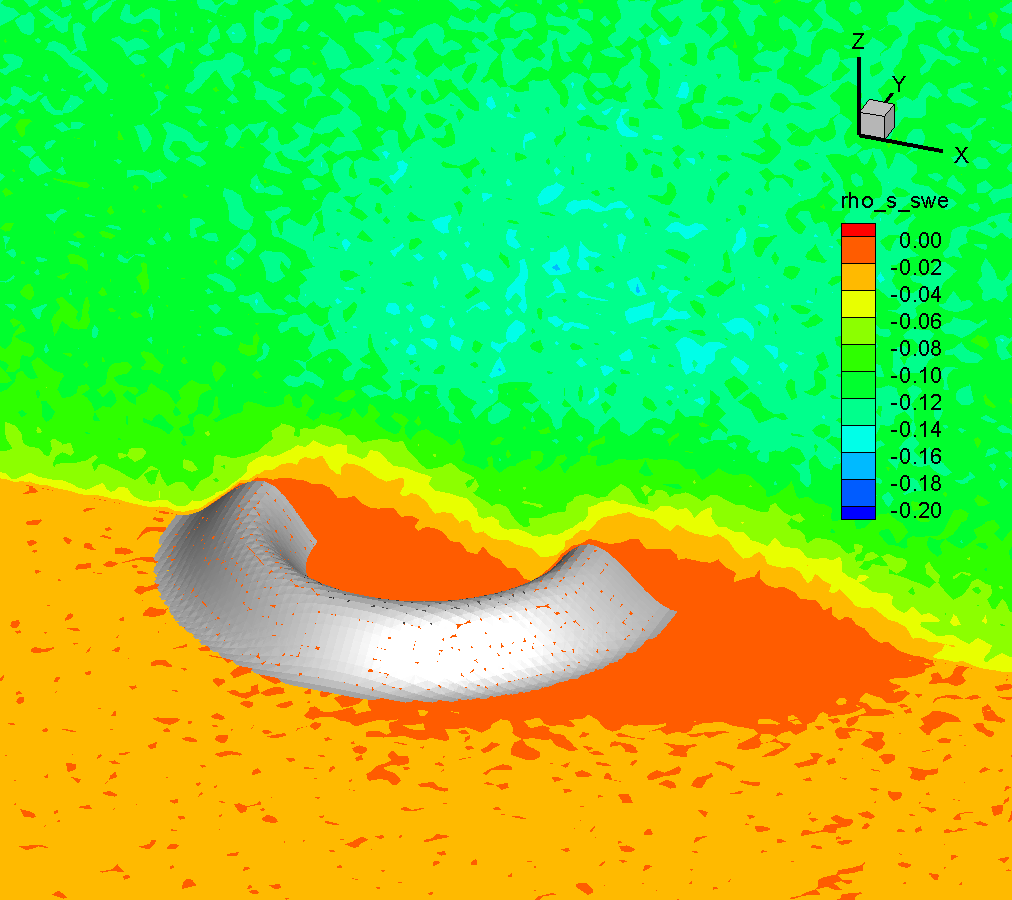}}
\subfigure[Photoelectron density contours.]%
{\includegraphics[width=0.45\textwidth]%
{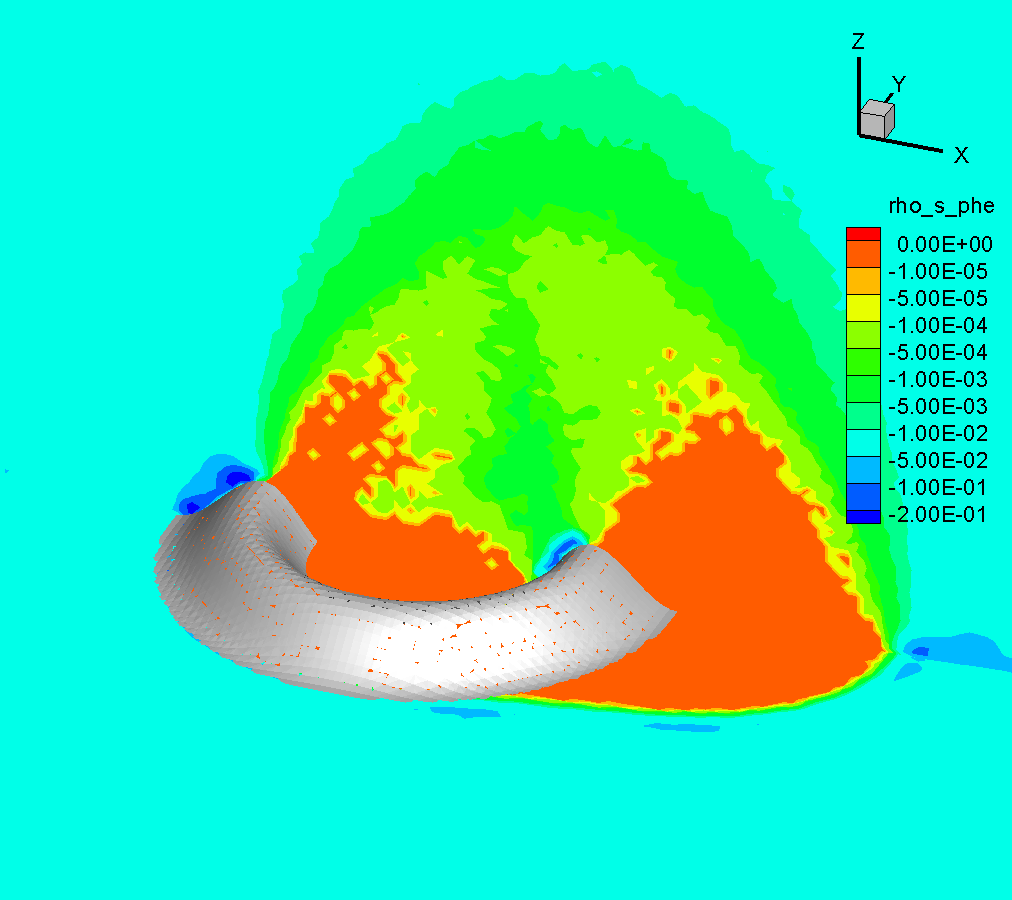}}
\subfigure[Total charge density contours.]%
{\includegraphics[width=0.45\textwidth]%
{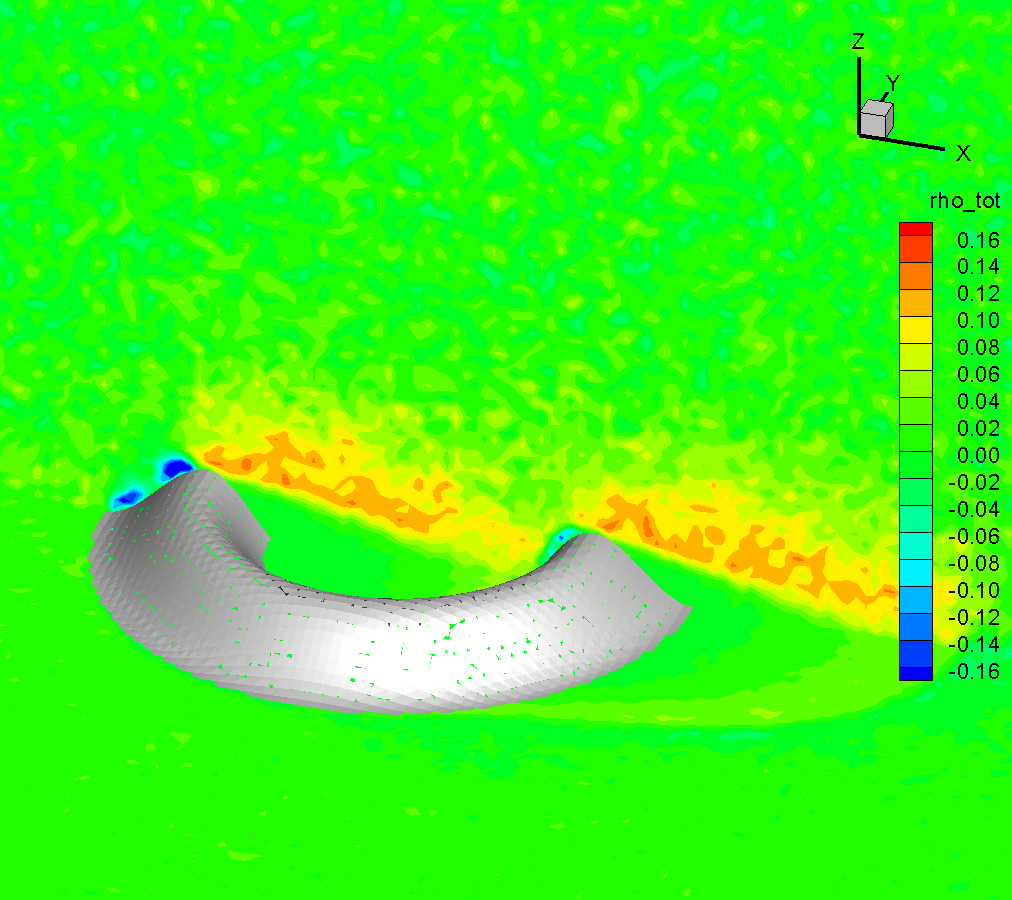}}
\end{subfigmatrix}
\caption{Normalized density contours.
For electrons, numerical values include a negative sign
indicating the negative charges.
The densities are normalized by 64 cm$^{-3}$
and the spatial dimensions are normalized by 1.38 m.}
\label{fig:lunar:densities}
\end{figure}

Figure \ref{fig:lunar:densities}
illustrate the density contours of solar wind ions, solar wind electrons,
photoelectrons, and total space charge near the crater.
The solar wind ion and electron density contours
clearly exhibit a localized plasma wake formed by the crater rim.
The photoelectron density contours clearly exhibit the lack of photoemission
in the shadow region.
The total space charge density contours show the non-neutral regions
associated with the wake caused by the crater rim.

\begin{figure}[ht!]
\centering
\begin{subfigmatrix}{2} 
\subfigure[Potential contours showing the differential charging near the lunar crater.]%
{\label{fig:d}\includegraphics[width=0.44\textwidth]%
{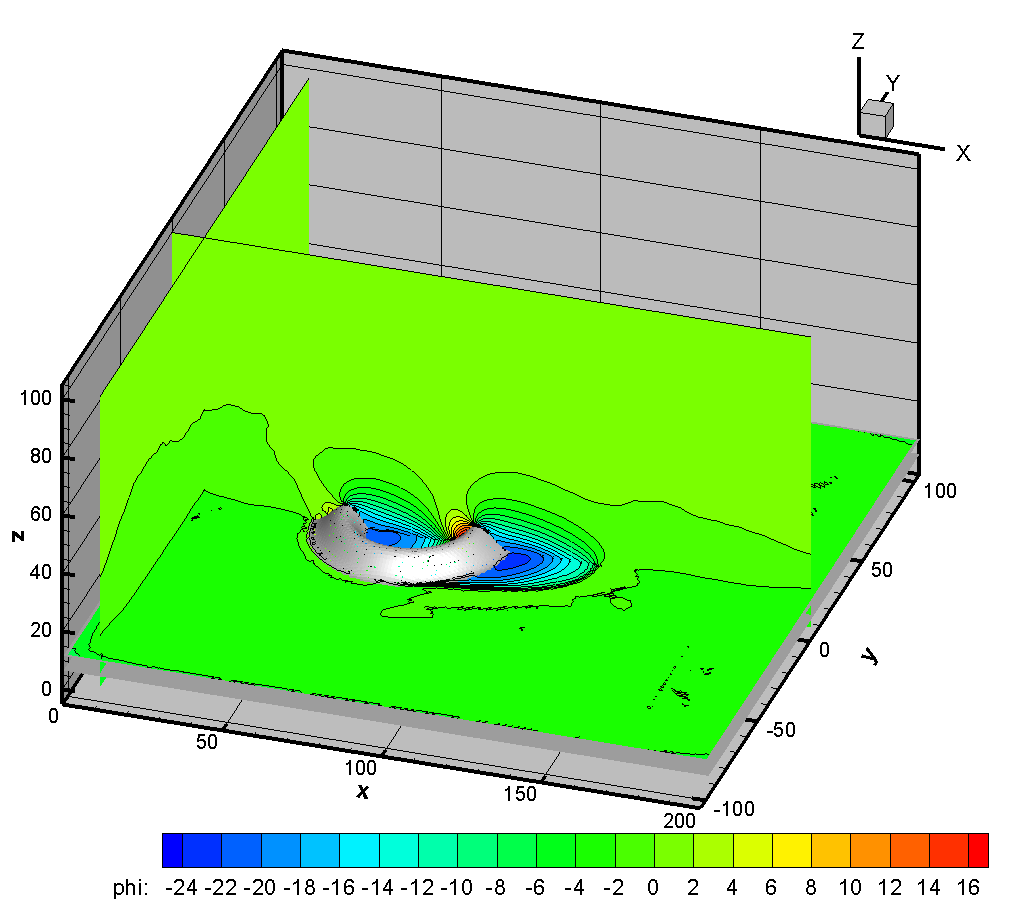}}
\subfigure[Zoom-in view of the potential contours near the lunar crater.]%
{\label{fig:dd}\includegraphics[width=0.44\textwidth]%
{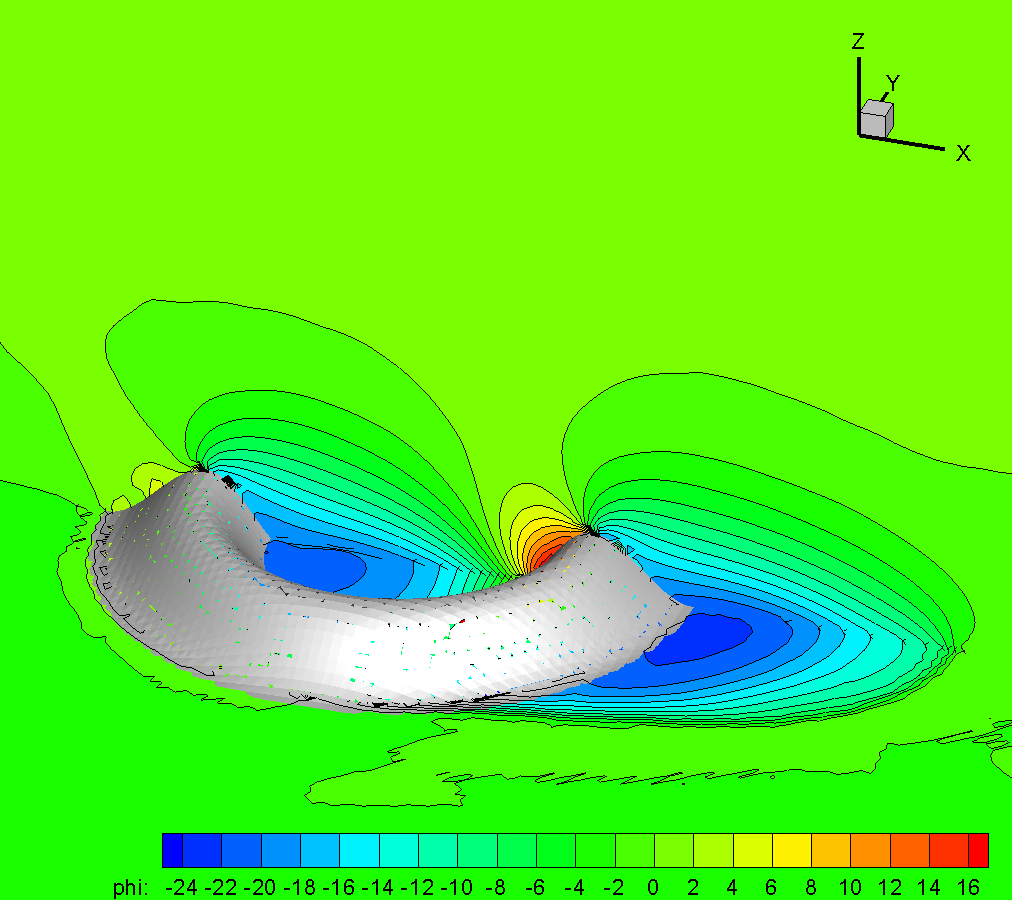}}
\end{subfigmatrix}
\caption{Potential contours of lunar crater charging.
The potential values are normalized by 2.2 V
and the spatial dimensions are normalized by 1.38 m.}
\label{fig:lunar:potential}
\end{figure}

Figure \ref{fig:lunar:potential}
illustrates the potential contours of the domain and near the crater.
It is shown,
for the average solar wind conditions considered here,
the surface potential in the sunlit region of the crater
is charged to about 16$\times$2.2 $\simeq$ 35 V
while the surface in the shadow region is charged to about
-24$\times$2.2 $\simeq$ -53 V.
It is noted as this length scale is on the order of tens of meters,
such differential surface charging will affect the lunar surface activities
for exploration missions, such as the risk of discharging/arcing
and horizontal/vertical transport of lofted charged lunar dusts.

\phantomsection
\section{Summary and Conclusion}
\label{sec:conclusion}

In this paper, we presented a most recently developed 3D
parallel immersed-finite-element particle-in-cell method,
namely, PIFE-PIC,
for kinetic particle simulations of plasma-material interactions
especially electrostatic surface charging.
PIFE-PIC is based on the serial non-homogeneous electrostatic IFE-PIC algorithm,
which was designed to handle complex interface conditions
associated with irregular geometries
while maintaining the computational speed Cartesian-mesh-based PIC.
3D domain decomposition is used in both field-solve and particle-push procedures of PIC
to distribute the computation among multiple processors.
A validation case of 3-D OML sheath of a dielectric sphere immersed
in a stationary plasma was carried out
and results agreed well with the analytic solution.
A series of strong scaling tests were performed to profile the parallel efficiency
for a problem of fixed size
which has 1 million PIC cells (5 million tetrahedral FE/IFE cells),
about 54 million particles,
and running 20,000 PIC steps
on the Foundry cluster at Missouri University of Science and Technology.
Parallel efficiency up to approximately 110\% superlinear speedup was achieved.


An application of PIFE-PIC to a larger problem, solar wind plasma charging at a lunar crater,
is presented to show the capability of PIFE-PIC for practical problems
of science and engineering interest.
The lunar crater charging simulation has 2 million PIC cells
(10 million tetrahedral FE/IFE cells),
about 520 million particles,
and running for 20,000 PIC steps.
The simulation finished in about 109 wall-clock hours
with domain decomposition of 8$\times$4$\times$4 = 128 MPI processes.
This demonstrates that PIFE-PIC can be utilized to carry out
realistic large-scale particle simulations of plasma-material interactions
routinely on supercomputers with distributed memory.

\clearpage
\phantomsection


\end{document}